\documentclass[aps,prb,12pt,preprint,notitlepage,superscriptaddress]{revtex4-1}
\usepackage{graphicx}
\usepackage{amsmath, amssymb}
\usepackage{siunitx}
\usepackage{bm}
\usepackage{pdfpages}

\makeatletter
\AtBeginDocument{\let\LS@rot\@undefined}
\makeatother

\bibpunct{[}{]}{,}{n}{}{}

\begin{document}


# Correlated Insulator Behaviour at Half-Filling in Magic Angle Graphene Superlattices

Yuan Cao[1], Valla Fatemi[1], Ahmet Demir[1], Shiang Fang[2], Spencer L. Tomarken[1], Jason Y. Luo[1], J. D. Sanchez-Yamagishi[2], K. Watanabe[3], T. Taniguchi[3], E. Kaxiras[2,4], R.C. Ashoori[1], P. Jarillo-Herrero[1,*]

[1] Department of Physics, Massachusetts Institute of Technology, Cambridge, Massachusetts 02139, USA
[2] Department of Physics, Harvard University, Cambridge, Massachusetts 02138, USA
[3] National Institute for Materials Science, Namiki 1-1, Tsukuba, Ibaraki 305-0044, Japan
[4] John A. Paulson School of Engineering and Applied Sciences, Harvard University, Cambridge, Massachusetts 02138, USA

**Van der Waals (vdW) heterostructures are an emergent class of metamaterials comprised of vertically stacked two-dimensional (2D) building blocks, which provide us with a vast tool set to engineer their properties on top of the already rich tunability of 2D materials. [1] One of the knobs, the twist angle between different layers, plays a crucial role in the ultimate electronic properties of a vdW heterostructure and does not have a direct analog in other systems such as MBE-grown semiconductor heterostructures. For small twist angles, the moiré pattern produced by the lattice misorientation creates a long-range modulation. So far, the study of the effect of twist angles in vdW heterostructures has been mostly concentrated in graphene/hexagonal boron nitride (h-BN) twisted structures, which exhibit relatively weak interlayer interaction due to the presence of a large bandgap in h-BN. [2-5] Here we show that when two graphene sheets are twisted by an angle close to the theoretically predicted 'magic angle', the resulting flat band structure near charge neutrality gives rise to a strongly-correlated electronic system. [6] These flat bands exhibit half-filling insulating phases at zero magnetic field, which we show to be a Mott-like insulator arising from electrons localized in the moiré superlattice. These unique properties of magic-angle twisted bilayer graphene (TwBLG) open up a new playground for exotic many-body quantum phases in a 2D platform made of pure carbon and without magnetic field. The easy accessibility of the flat bands, the electrical tunability, and the bandwidth tunability though twist angle may pave the way towards more exotic correlated systems, such as unconventional superconductors or quantum spin liquids.**

* pjarillo@mit.edu

Exotic quantum phenomena often occur in condensed matter and other systems with high density of states. Remarkable examples include superconductivity [7] and the fractional quantum hall effect [8]. One way of creating high density of states is to have flat bands with weak dispersion in momentum space, where the electron kinetic energy is set by the bandwidth $W$. When the Fermi level lies within the flat bands, Coulomb interactions ($U$) can then greatly exceed the kinetic energy of the electrons and drive the system into various strongly correlated phases ($\frac{U}{W} \gg 1$). [9-12] The study of such flat-band systems in bulk materials continues to be scientifically important, and the search for new flat-band systems, such as in kagome and Lieb lattices as well as in heavy fermion systems, is ongoing. [9-15]

Recent advances in 2D materials provide a new route for achieving flat bands. An inherent advantage in 2D is that the chemical potential of electrons can be continuously tuned via electric field effect without introducing extra disorder. In a twisted vdW heterostructure, the mismatch between two similar lattices generates a moiré pattern (Fig. 1b). This additional periodicity, which can have a length scale orders of magnitude larger than the underlying atomic lattices, has been shown to create a fractal energy spectrum in a strong magnetic field. [2-4] In twisted layers, the interlayer hybridization is modulated by the moiré pattern as well. As an example, the band structure of TwBLG can be tailored to generate band gaps and band curvatures otherwise absent in the graphene bands. [6, 16-20] Although the well-known building blocks for vdW heterostructures, such as graphene and transition metal dichalcogenides, do not have intrinsic flat bands at low energies, it has been predicted theoretically that flat bands may exist in TwBLG. [6, 17-19, 21] In this work, we demonstrate experimentally that when the twist angle of TwBLG is close to the theoretically predicted 'magic angle', the interlayer hybridization induces nearly-flat low-energy bands. This quenching of the quantum kinetic energy leads to a correlated insulating phase at half-filling of these flat bands, which points towards a Mott insulator in the localized flat bands.

To zeroth order, the low energy band structure of TwBLG can be considered as two sets of monolayer graphene Dirac cones rotated about the $\Gamma$ point by the twist angle $\theta$ (Fig. 1d). [6] The difference between the two $K$ (or $K'$) wave vectors gives rise to the mini Brillouin zone (MBZ), shown as a small hexagon, which is reciprocal to the moiré superlattice. The Dirac cones near the same valley mix through interlayer hybridization, while interactions between distant Dirac cones are exponentially suppressed. [6, 16] As a result, the valley itself remains (for all practical purposes) a good quantum number. Two experimentally verified consequences of this hybridization are energy gaps that open near the intersection of the Dirac cones and renormalization of the Fermi velocity $v_F = \frac{1}{\hbar}|\vec{\nabla}_{\vec{k}} E_{\vec{k}}|_{\vec{k}=K,K'}$ at the Dirac points. [16, 22-24]

2

The theoretically calculated 'magic angles' $\theta_{magic}^{(i)}$, $i = 1, 2, ...$ are a series of twist angles at which the Fermi velocity at the Dirac points becomes zero.[6] The resulting low-energy bands near these twist angles are confined to less than about 10 meV. These phenomena can be qualitatively understood from the competition between the kinetic energy and interlayer hybridization energy (Fig. 1e-g). Intuitively, when the hybridization energy $2w$ is comparable or larger than $\hbar v_0 k_\theta$ where $v_0 = 10^6$ m/s is the Fermi velocity of graphene and $k_\theta \approx K\theta$ is the momentum displacement of the Dirac cones, the lower of the hybridized states is pushed to and crosses zero energy. A detailed treatment gives the first magic angle $\theta_{magic}^{(1)} = \frac{\sqrt{3}w}{\hbar v_0 K}$, which is approximately 1.1°.[6] Fig. 1c shows an *ab initio* tight-binding calculation of the band structure for $\theta = 1.08°$.[18] The labeled flat bands have a bandwidth of 12 meV for the $E > 0$ branch and 2 meV for the $E < 0$ branch (strain effects can somewhat modify these values, but not drastically).[25] From a band theory point of view, the flat bands should have localized wave function profiles in real space. Fig. 1h shows the calculated local density of states for the flat bands. The wave functions are indeed highly concentrated in the regions with AA stacking, while small but finite amplitudes on the AB and BA regions connect the AA regions and endow a weak dispersion to the bands.[21] A brief discussion about the topological structure of the bands near the first magic angle is given in the supplementary information.[26]

For the experiment, we fabricated high-quality encapsulated TwBLG devices with twist angle precisely controlled to 0.1~0.2° accuracy using a previously developed 'tear and stack' technique.[16,27] We have measured four devices that have twist angles near the first magic angle $\theta_{magic}^{(1)} \approx 1.1°$. Fig. 2a shows the low-temperature two-probe conductance of device D1 as a function of carrier density $n$. Near $n = \pm n_s = \pm 2.7 \times 10^{12}$ cm$^{-2}$ (4e$^-$ per moiré unit cell for the twist angle $\theta = 1.08°$), the conductance is zero over a wide range of densities. These insulating states have been previously understood as hybridization induced band gaps above and below the lowest energy superlattice bands, and will be hereafter referred to as the 'superlattice gaps'.[16] The measured thermal activation gaps are about 40 meV.[16,26] The twist angle can be estimated from the density required to reach the superlattice gaps, which we find to be $\theta = 1.1 \pm 0.1°$ for all of the devices reported here.[26]

A new pair of insulating phases occurs for a narrower density range near half of the superlattice density $n = \pm \frac{n_s}{2} = \pm 1.4 \times 10^{12}$ cm$^{-2}$ (2e$^-$ per moiré unit cell). These insulating states have a much smaller energy scale. Note that this behaviour is markedly different from all other zero-field insulating behaviors previously reported in the recent

3

literature (which occur at either $\pm n_s$ or at an integer multiple of $\pm n_s$),[16, 20] and we shall refer to them as half-filling insulating phases (HFIPs). These states are observed at roughly the same density for all four devices (Fig. 2a inset). Fig. 2b and Fig. 2c-d show the conductance of the HFIPs in device D1 at different temperatures. Above 4 K, the system behaves as a metal, exhibiting decreasing conductance with increasing temperature. A metal-insulator transition occurs near 4 K. The conductance drops significantly from 4 K to 0.3 K, with the minimum value decreasing by 1.5 orders of magnitude. An Arrhenius fit gives a thermal activation gap of ~0.3 meV for the HFIPs, two orders of magnitude smaller than those of the superlattice gaps. At the lowest temperatures the system may be limited by conduction through charge puddles, resulting in deviation from the Arrhenius fit.

To confirm the existence of the HFIPs, we performed capacitance measurements on device D2 using an ac low-temperature capacitance bridge.[26, 28] The real and imaginary components of the ac measurement give information about the change in capacitance and the loss tangent of the device, respectively. The latter signal is tied to the dissipation in the device due to its resistance.[26] Device D2 exhibits a reduction in capacitance and strong enhancement of dissipation at $\pm\frac{n_s}{2}$ as shown in Fig. 3a, in agreement with an insulating phase that results from the suppression in density of states. The insulating state at $-\frac{n_s}{2}$ is weaker and only visible in the dissipation data. The observation of capacitance reduction (i.e. suppression of density of states) for only the n-side HFIP in this device may be due to an asymmetric band structure and/or device quality. The reduction (enhancement) in capacitance (dissipation) vanishes when the device is warmed up from 0.3 K to about 2 K, consistent with the behavior observed in transport measurements.

The emergence of HFIPs is not expected in a single-particle picture and appears to be correlated with the narrow bandwidth near the first magic angle. In our experiment, several separate pieces of evidence support the presence of flat bands. First, we measured the temperature dependence of the amplitude of Shubnikov-de Haas (SdH) oscillations in device D1, from which we extracted the electron effective mass $m^*$ (Fig. 3b, see SI for details).[26] For a Dirac spectrum with eight-fold degeneracy (spin, valley, and layer), one would expect that $m^* = \sqrt{\frac{h^2 n}{8\pi v_F^2}}$ which scales as $v_F^{-1}$. The large measured $m^*$ near charge neutrality in device D1 indicates a 25-fold reduction in $v_F$ ($v_F$=4×10⁴ m/s, compared to 10⁶ m/s in monolayer graphene). This striking reduction of the Fermi velocity is an expected characteristic of the flat bands. Furthermore, we analyzed the capacitance data of device D2 near the Dirac point (Fig. 3a), finding that a Fermi velocity reduced to about $0.15 v_0$ is necessary for a good fit to the data (see Fig. S4 in SI).[26] Finally, another direct manifestation of such flat bands is the flattening of the

4

conductance minimum at charge neutrality above a temperature of 40 K ($kT$=3.5 meV), as seen in Fig. 3c. While the conductance minimum in monolayer graphene can be clearly observed even near room temperature,[29-31] it is smeared out in magic-angle TwBLG when the thermal energy $kT$ becomes comparable to $\frac{v_F k_\theta}{2} \approx$ 4 meV, the energy scale spanning the Dirac-like portion of the band (see Fig. 1c).

Due to the localized nature of the electrons, a plausible explanation for the gapped behavior at half-filling is the formation of a Mott-like insulator driven by Coulomb interaction between electrons.[32, 33] To this end, we can consider a Hubbard model on a triangular lattice, where each site corresponds to a localized region with AA stacking in the moiré pattern (Fig. 1i). Fig. 3d shows the numerically calculated bandwidth of the $E>0$ branch of the low-energy bands for $0.04<\theta<2°$ using a continuum model of TwBLG.[6] The bandwidth $W$ is strongly suppressed near the magic angles. The on-site Coulomb energy $U$ of each site is estimated as $\frac{e^2}{4\pi\varepsilon d}$, in which $d$ is the effective linear dimension of each site (with the same length scale as the moiré period) and $\varepsilon$ is the effective dielectric constant including screening. Absorbing $\varepsilon$ and the dependence of $d$ on twist angle into a single constant $\kappa$, we can write $U = \frac{e^2 \theta}{4\pi\varepsilon_0 \kappa a}$ ($a$=0.246 nm is the lattice constant of monolayer graphene). In Fig. 3d we plot the on-site energy $U$ versus $\theta$ for $\kappa$=4-20. As a reference, $\kappa$=4 if one assumes $\varepsilon = 10\varepsilon_0$ and $d$ equals 40% of the moiré wavelength. For a range of possible values of $\kappa$ it is therefore reasonable that $U/W > 1$ occurs near the magic angles and results in half-filling Mott-like gaps.[32] The realistic scenario is however much more complicated than these simplistic estimates and would require detailed theoretical analysis of the interactions responsible for the correlated gaps.

The SdH oscillation frequency, shown on the right-hand axis of Fig. 3b, also supports the existence of Mott-like correlated gaps at half-filling. Near the charge neutrality point, the oscillation frequency closely follows $f_{SdH} = \frac{\phi_0 |n|}{N}$ where $\phi_0 = \frac{h}{e}$ is the flux quantum and $N = 4$ indicates the spin and valley degeneracies. Surprisingly, at $|n| > \frac{n_s}{2}$ we observe oscillation frequencies that corresponds to straight lines $f_{SdH} = \frac{\phi_0 (|n| - \frac{n_s}{2})}{N}$ in which $N$ has a reduced value of 2. Moreover, these lines extrapolate to zero exactly at the densities of the HFIPs $n = \pm\frac{n_s}{2}$. These oscillations point towards small Fermi pockets that result from doping the HFIP states, which might originate from charged quasiparticles near a Mott-like insulator phase.[34] The halved degeneracy of the Fermi pockets might be related to the spin-charge separation predicted in a Mott insulator.[34]

These results are also supported by Hall measurements at 0.3 K presented in the supplementary information, that show a "resetting" of the Hall densities when the system is electrostatically doped beyond the Mott-like states. [26]

The HFIPs at $\pm\frac{n_s}{2}$ are suppressed by the application of a magnetic field. Figures 4a-b shows that both insulating phases start to conduct at a perpendicular field of $B$=4 T and recover normal conductance by $B$=8 T. A similar effect is observed for in-plane magnetic field. [26] The insensitivity to field orientation suggests the suppression of the HFIPs is due to a Zeeman effect rather than an orbital effect, as the latter would be affected only by the perpendicular component of the magnetic field. For an effective g-factor $g = 2$ due to electron spin, the Zeeman energy needed to suppress the HFIPs is approximately $g\mu_B B$=0.5 meV, on the same order as the thermal excitation energy scale.

Our data point towards the presence of a spin-singlet Mott-like insulator ground state at half-filling and zero magnetic field (Fig. 4e). The application of an external magnetic field can polarize the excitations in the spectrum of the correlated states according to their spin. When the Zeeman energy exceeds the charge gap, charge conduction can therefore occur (Fig. 4f). In a typical Mott insulator, the ground state usually has an antiferromagnetic spin ordering below the Néel temperature. On a triangular lattice, however, the frustration prevents the fully anti-parallel alignment of adjacent spins. Possible ordering schemes include 120° Néel order and rotational symmetry breaking stripe order. [35] It is yet unclear whether the spin-singlet ground state in TwBLG is fulfilled by any of the above ordering schemes or simply disordered at low temperatures. In the HFIPs of TwBLG it is also possible that the ordering, if any, occurs in conjunction with the valley degree of freedom. Therefore, a complete theoretical treatment of this problem should at least involve considering a two-band Hubbard model on a triangular lattice.

We also comment on other competing mechanisms for creating a half-filled insulating state in a system with flat bands. Among the possibilities, charge density waves (CDW) in 2D are often stabilized by Fermi surface nesting, which can in principle occur near the half-filling of a 2D Brillouin zone. [36] However, the nesting is not sufficient to fully gap out the entire Fermi surface to achieve an insulating state. In order to create a global gap at half-filling, at least a doubling of the unit cell would be necessary, which could be created by a commensurate CDW or lattice relaxation due to strain. Scanning tunneling experiments conducted at temperatures below 1 K may be able to differentiate such mechanisms.

In summary, our work demonstrates that graphene can be transformed through van der Waals engineering into a flat-band system in which correlation effects play a fundamental

role. Through its easy gate tunability, magic-angle TwBLG provides a novel playground for studying the Mott transition, which may provide insight into strongly-correlated materials, notably high-temperature superconductivity. The richness of the combined spin and valley degrees of freedom on a triangular lattice could also give rise to other exotic quantum phases such as quantum spin liquids.[37]


## Methods

**Sample Preparation**
Device D1, D2 and D4 are fabricated using a modified 'tear & stack' technique detailed in previous works.[16, 27] Monolayer graphene and hexagonal boron nitride (h-BN, 10 to 30 nm thick) are exfoliated on $SiO_2$/Si chips and examined with optical microscopy and atomic force microscopy. We use a Poly(Bisphenol A carbonate) (PC)/Polydimethylsiloxane (PDMS) stack on a glass slide mounted on a micro-positioning stage to first pick up h-BN flake at 90˚C. Then we use the van der Waals force between h-BN and graphene to tear a graphene flake at room temperature. The separated graphene pieces are manually rotated by an angle $\theta$ about 0.2~0.3° larger than the desired twist angle and stacked together again, resulting in a precisely controlled TwBLG structure. The TwBLG is then encapsulated by picking up another h-BN flake on the bottom, and the entire stack released onto a metal gate at 160˚C. The final device geometry is defined by electron-beam lithography and reactive ion etching. Device D3 is fabricated using a slightly different procedure, where independent graphene flakes are stacked together. The edges of graphene flakes are aligned under optical microscope to obtain small twist angles.

**Measurements**
Transport measurements are performed using a standard low frequency lock-in amplifier with excitation frequency of 10~20 Hz and excitation voltage of 100 µV, in a He-3 cryostat. The current flowing through the device is amplified by a current pre-amplifier and then measured by the lock-in amplifier.

Capacitance was measured using a low-temperature balanced capacitance bridge.[28] The reference capacitance $C_{\text{ref}}$ used in our experiment is approximately 40 fF, and the device geometrical capacitance is approximately 7 fF. The ac excitation voltage used in our measurements is 3 mV at $f$ =150 kHz.

**Supplementary Information**
Supplementary Information is available in the online version of the paper.

**Data Availability**

The data that support the findings of this study are available from the corresponding authors upon reasonable request.

**Author Contributions**
Y.C., J.Y.L., J.D.S-Y fabricated the devices and performed transport measurements. Y.C., V.F. performed data analysis. P.J.H. supervised the project. S.F. and E.K. provided numerical calculations. S.L.T., A.D. and R.C.A. measured capacitance data. K.W. and T.T. provided h-BN devices. Y.C., V.F., and P.J.H. wrote the paper with input from all authors.

**Author Information**
The authors declare no competing financial interest.

**Figure Captions**

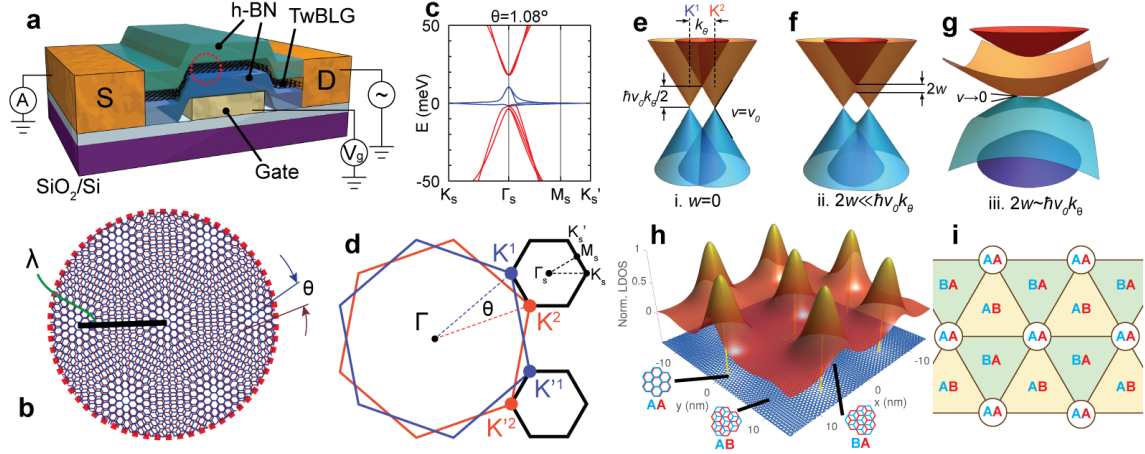

Figure 1. (a) Schematic of the twisted bilayer graphene (TwBLG) devices. The TwBLG is encapsulated in h-BN flakes of 10~30 nm thickness. The conductance is measured with a voltage bias of 100 µV while varying the local bottom gate voltage. (b) The moiré pattern as seen in TwBLG. The moiré wavelength $\lambda = \frac{a}{2\sin\frac{\theta}{2}}$ where $a = 0.246$ nm is the lattice constant of graphene and $\theta$ is the twist angle. (c) The band structure of $\theta=1.08°$ TwBLG calculated with an *ab initio* tight-binding method. The bands shown in blue are the flat bands under study in this work. (d) The MBZ is constructed from the difference between the two $K$ ($K'$) wave vectors from the two layers. Hybridization occurs between Dirac cones within each valley, while intervalley processes are strongly suppressed. (e-g) Illustration of the effect of interlayer hybridization for (e) $w = 0$, (f) $2w \ll \hbar v_0 k_\theta$ and (g) $2w \sim \hbar v_0 k_\theta$. (h) Calculated local density of states (LDOS) for the flat bands with E > 0 at $\theta=1.08°$. The electron density is strongly concentrated at the regions with A-A stacking order, while mostly depleted at A-B and B-A stacked regions. (i) Top view of a simplified model of the stacking order.

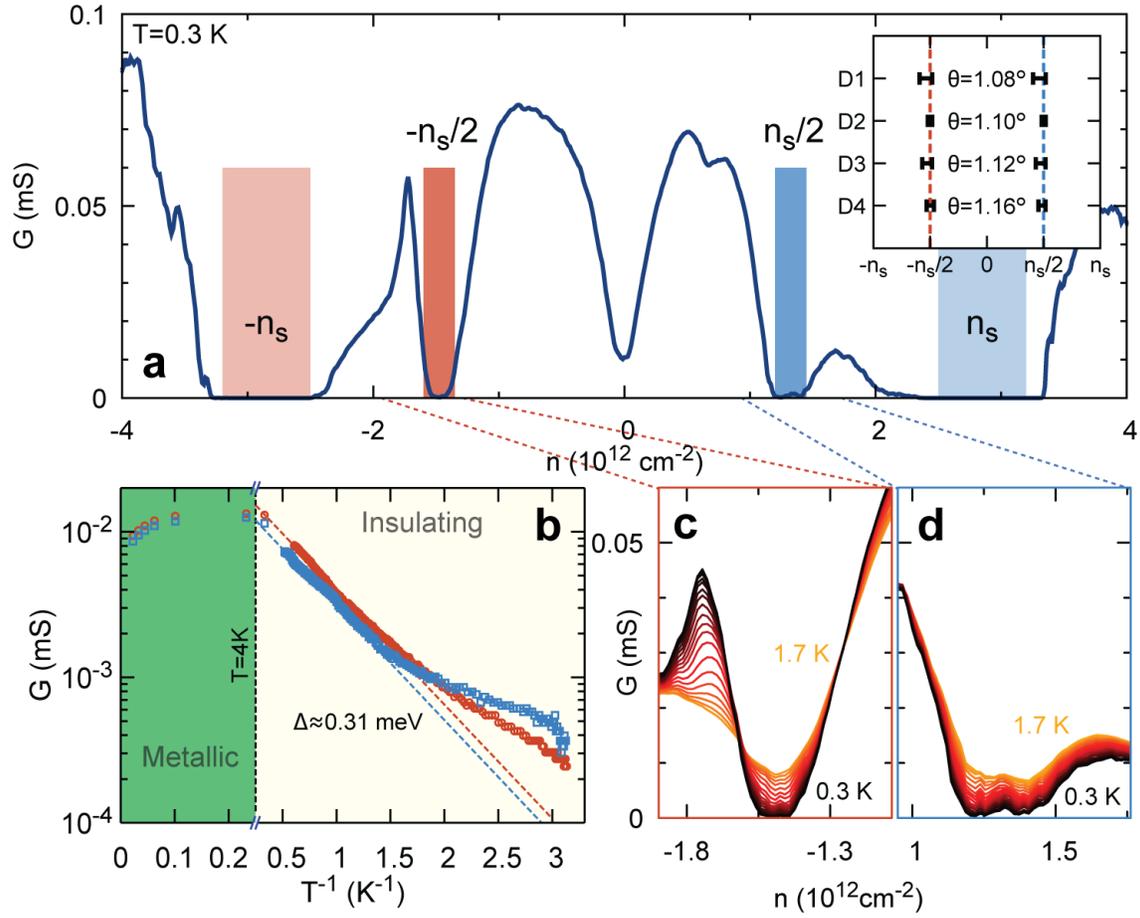

Figure 2. (a) Measured conductance of magic-angle TwBLG device D1 with $\theta=1.08°$. Dirac point is located at $n = 0$. The lighter shaded regions are superlattice gaps at carrier density $n = \pm n_s = \pm 2.7\times 10^{12}$ cm$^{-2}$. The darker shaded regions denote HFIPs at $\pm\frac{n_s}{2}$.

Inset shows the density locations of half-filling insulating phases (HFIPs) in four different devices. The method for obtaining the error bars is explained in the supplementary information.[26] (b) Minimum conductance values in the two HFIPs in device D1, labeled by corresponding colors as defined in (a) and (c-d). The dashed lines are fits of the formula $\exp(-\Delta/2kT)$ to the data, where $\Delta \approx 0.31$ meV is the thermal activation gap. (c-d) Temperature dependent conductance of D1 from 0.3~1.7 K near the (c) p-side and (d) n-side HFIPs.

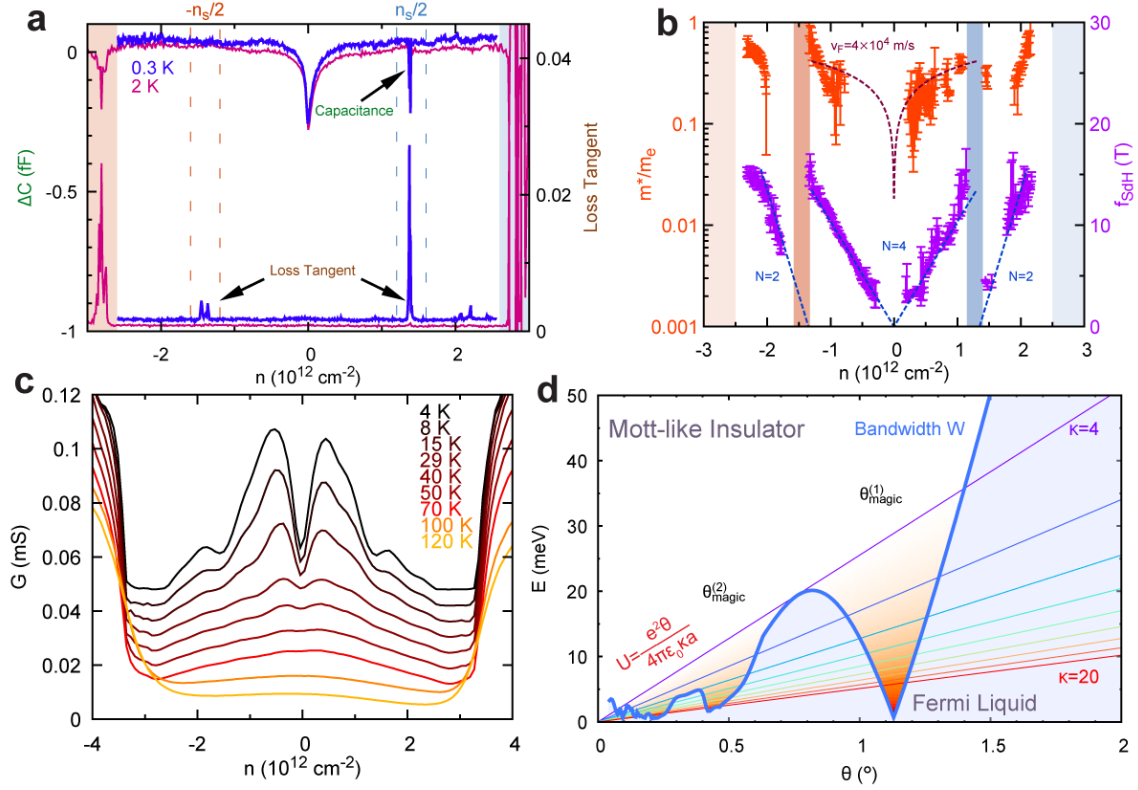

Figure 3. (a) Capacitance measurements of device D2 at 0.3 K (blue trace) and 2 K (pink trace). The change in the measured capacitance (upper traces), $\Delta C$, is plotted on the left axis, and the loss tangent (lower traces) is shown on the right axis. For density corresponding to half-filling $\pm\frac{n_s}{2}$, reduction in $\Delta C$ (on p-side only) and enhancement in loss tangent (on both sides) are observed (0.3 K data). These effects disappear in the 2 K measurements. (b) The effective mass $m^*$ and oscillation frequency $f_{SdH}$ as extracted from temperature-dependent SdH oscillations. The fitting curves are $m^* = \sqrt{\frac{h^2 n}{8\pi v_F^2}}$, assuming a uniform Fermi velocity $v_F$. For magic-angle device D1 the estimated Fermi velocity $v_F=4\times10^4$ m/s is 25-times reduced from that in pristine graphene, $v_0=1\times10^6$ m/s. The measured oscillation frequencies point towards the existence of small Fermi pockets that start from the HFIPs with one half the degeneracy of the main Fermi surface of the Dirac points. Shaded regions at half-filling and full-filling correspond to the shaded rectangles in Fig. 2a. (c) Gate-dependence of the conductance of D1 at different temperatures from 4.5 K to 120 K. The curves are vertically shifted for clarity. (d) The comparison between the bandwidth $W$ for the E > 0 flat band branch in TwBLG and the on-site energy $U$ for different twist angles. Near the magic angles $\theta_{magic}^{(i)} \approx 1.1°, 0.5°, ...$, $U > W$ is satisfied for a range of possible $\kappa$ (defined in the main text), and the system can be driven into a Mott-like insulator.

13

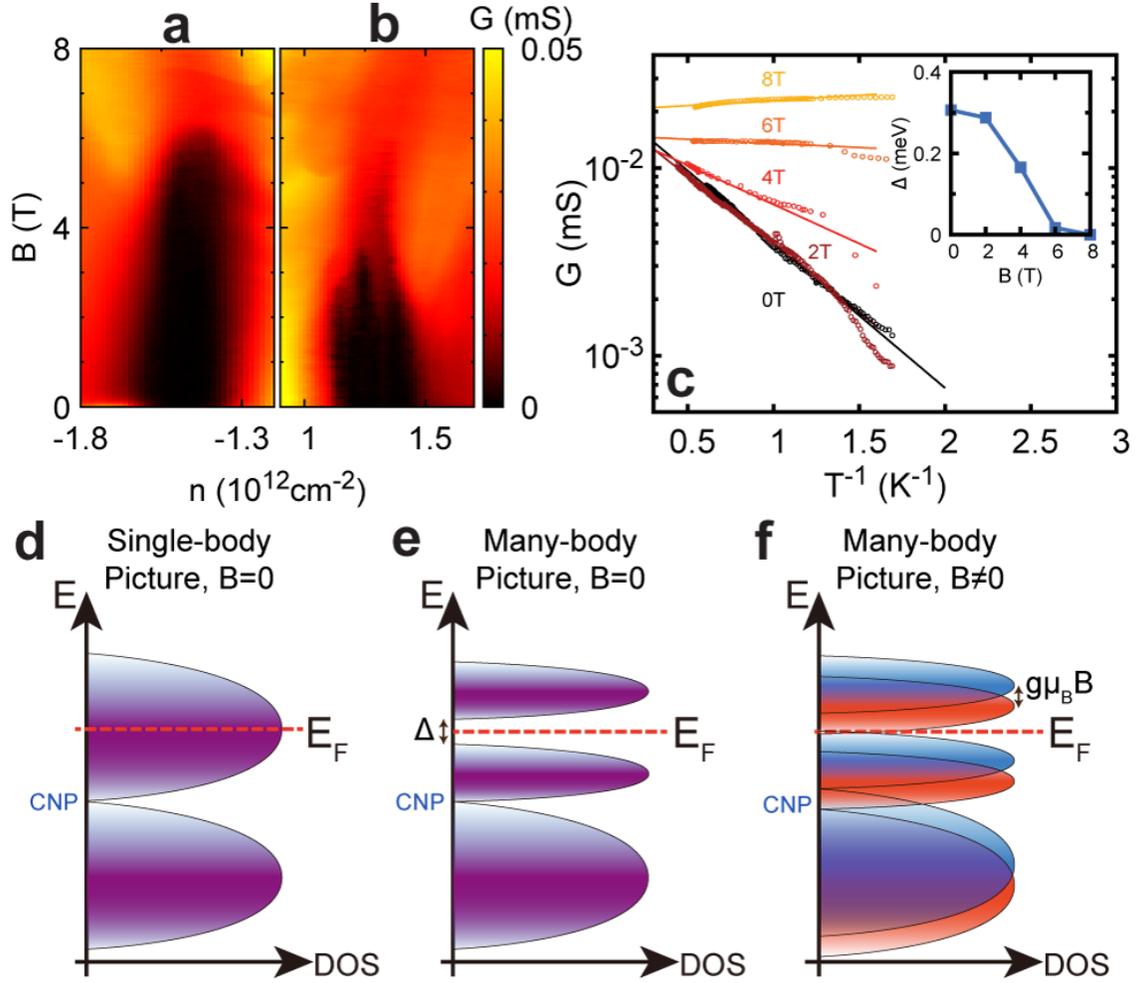

Figure 4. $B_\perp$ dependence of the conductance of the HFIPs for D1 on (a) p-side and (b) n-side. The white line cuts show the conductance at the labeled fields. (c) Arrhenius plot of the conductance of the p-side HFIP at different magnetic fields. The inset shows the thermal activation gap extracted from fitting the data of the main plot with $\exp(-\Delta/2kT)$. (d-f) Schematic density of states (DOS) pictures. The single-particle flat-bands (both E>0 and E<0 bands are shown, with $E_F$ in the E>0 band, i.e. n-doping) in (d) is split into upper and lower Hubbard bands by interactions (e). This occurs when $E_F$ is at half-filling of the upper band. Upon applying a Zeeman field, the excitations can be further polarized and close the charge gap when the Zeeman energy is comparable to the gap (f). Purple shading denotes a spin-degenerate band, while blue and red shading denote spin-up and spin-down bands respectively. CNP abbreviates for charge neutrality point. The shape of the DOS drawn here is purely illustrative and does not represent the actual DOS profile (see supplementary information for a numerical result).[26]



\title{Supplementary Information}

%\author{Y. Cao}
%\author{V. Fatemi, \emph{et. al.}}
\date{\today}
\maketitle

\tableofcontents
\newpage

\section{Sample Fabrication and Measurement}

Device D1, D2 and D4 are fabricated using a modified 'tear \& stack` technique detailed in our previous work \cite{cao2016} and Ref. \cite{kim2016, kim2017}. Briefly speaking, monolayer graphene and hexagonal boron nitride (h-BN, \SIrange{10}{30}{\nano\meter} thick) are exfoliated on SiO\textsubscript{2}/Si chips and examined with optical microscopy and atomic force microscopy. We use PC/PDMS stack on a glass slide mounted on a micro-positioning stage to first pick up the h-BN flake. Then we use the van der Waals force between h-BN and graphene to tear a graphene flake. The separated graphene pieces are manually rotated by a twist angle $\theta$ and stacked on each other, resulting in the desired TwBLG structure. The resulting stack is encapsulated with another h-BN flake at the bottom and put onto a metal gate. The final device geometry is defined by electron-beam lithography and reactive ion etching. Electrical connections to the TwBLG are made by one-dimensional edge contacts \cite{wang2013}. Device D3 is fabricated using a slightly different procedure, where independent graphene flakes are stacked together. The edges of graphene flakes are aligned under an optical microscope to obtain small twist angles.

All transport measurements are performed using standard lock-in techniques with excitation frequency between \SIrange{10}{20}{\hertz} and excitation voltage $V_\mathrm{drive} = \SI{100}{\micro\volt}$. The current flowing through the sample is amplified by a current pre-amplifier and then measured by the lock-in amplifier.

\section{Supplemental Transport Data}

\subsection{Full Temperature Dependence of Device D1}

\begin{figure}[!htb]
	\includegraphics[width=\textwidth]{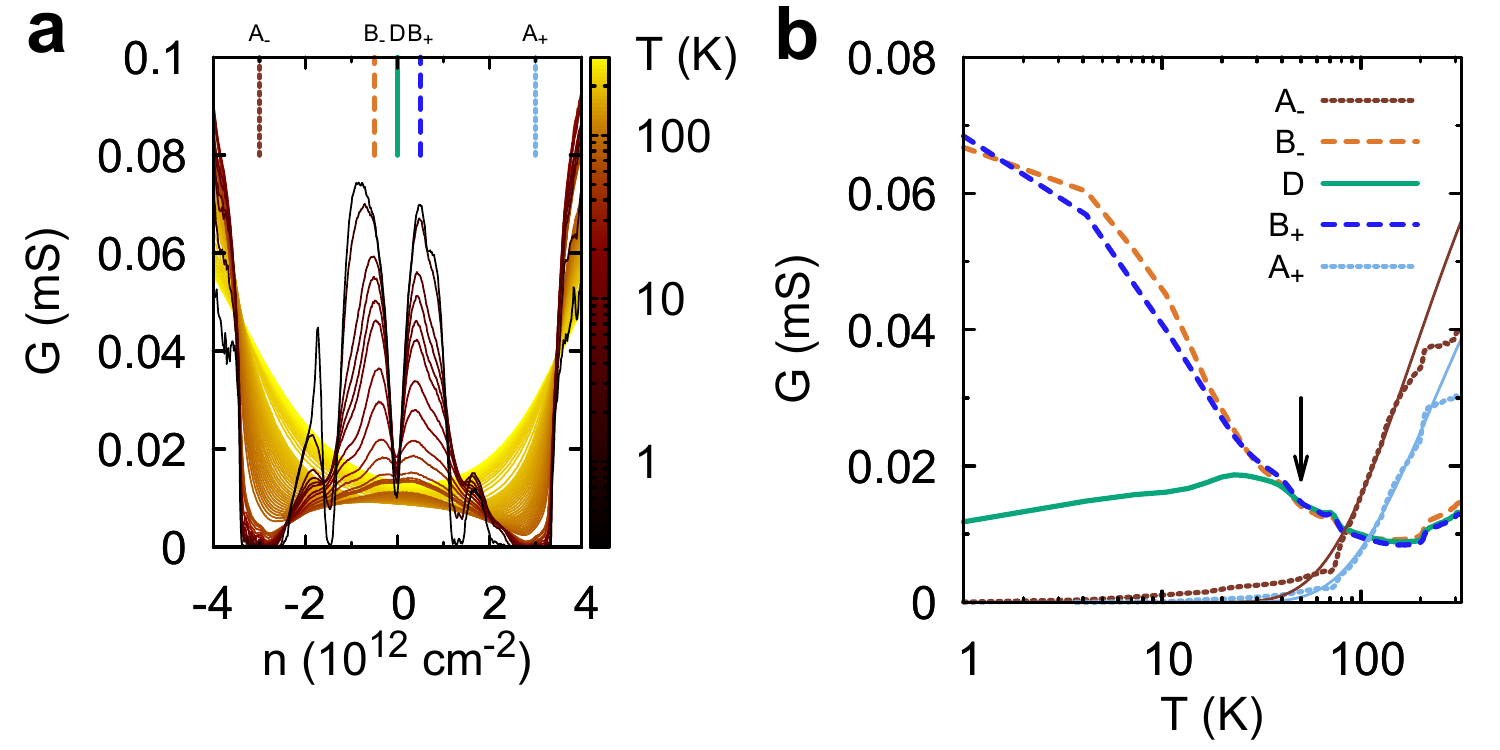}
	\caption{\label{fig:fulltemp} (a) Temperature dependence of conductance of device D1 from \SI{0.3}{\kelvin} to \SI{300}{\kelvin}. (b) The conductance versus temperature at five characteristic carrier densities labeled A\textsubscript{$\pm$} (superlattice gaps), B\textsubscript{$\pm$} (above and below the Dirac point) and D (the Dirac point) in (a). The arrow denotes the temperature above which the conductances at B\textsubscript{$\pm$} merge with $D$. The solid lines accompanying $A_\pm$ are Arrhenius fit to the data.}
\end{figure}

Fig. \ref{fig:fulltemp} shows the full temperature dependence of the conductance of device D1 from \SI{0.3}{\kelvin} to \SI{300}{\kelvin}. The metallic behavior at all densities from \SIrange{4}{100}{\kelvin} except for the superlattice gaps (A\textsubscript{$\pm$}) are clearly seen in Fig. \ref{fig:fulltemp}(b).

Lines labeled B\textsubscript{-}, D, B\textsubscript{+} correspond to below, at, and above the Dirac point respectively. The conductance values at these densities completely merge at about \SI{50}{\kelvin}, indicative of the narrow bandwidth as described in the main text. 

The thermal activation gaps of the superlattice insulating states at A\textsubscript{$\pm$} can be obtained by fitting the temperature dependence of the conductance at these densities. Detailed discussion about the superlattice gaps in non-magic-angle devices are published in Ref. \cite{cao2016}. The fit to Arrhenius formula $\exp(-\Delta/2kT)$ yields $\Delta=\SI{32}{\milli\electronvolt}$ for the A\textsubscript{-} gap and \SI{40}{\milli\electronvolt} for the A\textsubscript{+} gap. For comparison, the same gaps measured in $\theta=\SI{1.8}{\degree}$ TwBLG are slightly larger at \SI{50}{\milli\electronvolt} and \SI{60}{\milli\electronvolt} for the gaps at negative and positive densities respectively \cite{cao2016}.

\subsection{Magnetotransport in Devices D1 and D3}
\label{sec:ll}

\begin{figure}[!htb]
	\includegraphics[width=\textwidth]{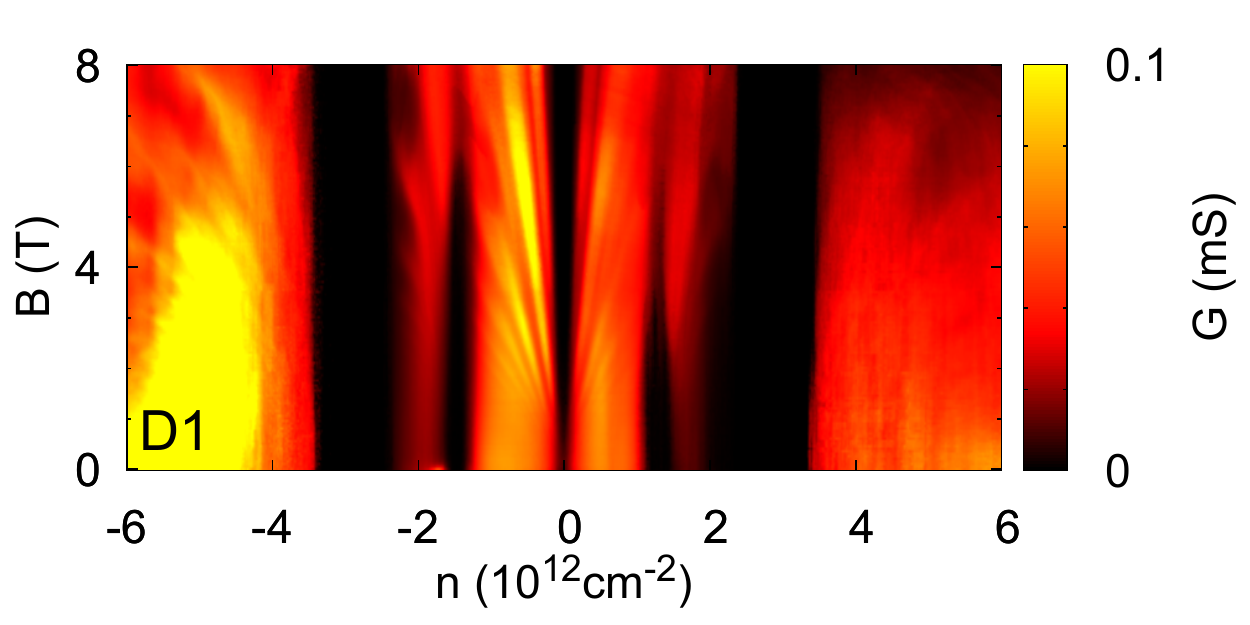}
	\includegraphics[width=\textwidth]{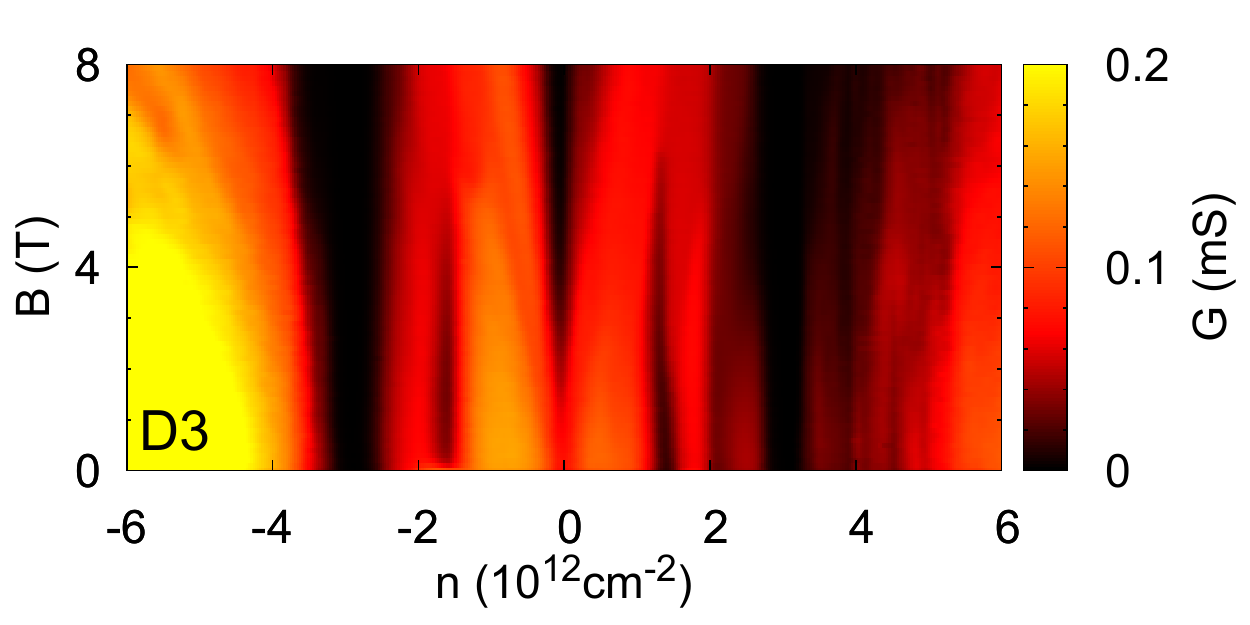}
	\caption{\label{fig:mt13}Magneto-conductance of two different samples D1 (also shown in main text) and D3. Both measurements are taken at temperatures $T\le\SI{0.3}{\kelvin}$.}
\end{figure}

Magnetoconductance data for device D1 and D3 are plotted in Fig. \ref{fig:mt13}. Magnetic field is applied perpendicular to the sample plane.

As in any clean metallic electronic system, Landau levels become visible above a certain magnetic field, and the conductance shows an oscillatory behavior periodic in $1/B$. Here we observe the onset of quantum oscillation at \SI{1.3}{\tesla} in D1 and \SI{2}{\tesla} in D3 (much weaker). The degeneracy of each Landau level is given by $n_d = B/\phi_0$ where $\phi_0=h/e$ is the flux quantum. Therefore the filling factor $\nu=n/n_d=n\phi_0/B$ can be directly read out from the slope of each Landau level. The observed Landau levels near the charge neutrality point in magic-angle samples have a filling factor sequence of $\pm(4, 8, 12, \ldots)$, in contrast to the $\pm(2, 6, 10, \ldots)$ sequence seen in monolayer graphene \cite{novoselov2005} or $\pm(4, 12, 20, \ldots)$ observed in non-magic-angle TwBLG (which is just twice that of monolayer graphene) \cite{cao2016, kim2016_2}. We find this sequence in all measured TwBLG devices with $\theta < \SI{1.3}{\degree}$. The same sequence is also observed in Ref. \cite{kim2017} at \SI{0.97}{\degree}. 

%It is consistent with a parabolic band touching at zero energy, similar to Bernal-stacked bilayer graphene. 

%However, it is confirmed both theoretically and experimentally that at larger twist angles TwBLG retain a massless Dirac-like band dispersion at low energies \cite{bistritzer2011, jdsy2012, cao2016}. This discrepancy leads naturally to the question of how the band structure and its associated topological properties cross over from Dirac-like to parabolic as the twist angle approaches the first magic angle. We will address this question in Sec. \ref{sec:band}. The conclusion is that at the magic angle where the Fermi velocity approaches zero, the low-energy dispersion is indeed analogous to Bernal-stack bilayer graphene with a Berry phase of $-2\pi$ (original Dirac point has $\pi$), explaining the unexpected Landau level sequence observed here. 

The observation of superlattice gaps near $n=\pm\SI{2.7e12}{\per\centi\meter\squared}$ and half-filling insulating phases (HFIPs) near $n=\pm\SI{1.4e12}{\per\centi\meter\squared}$ is consistent across both devices. The HFIPs are also suppressed at a similar magnetic field, between 4 and \SI{6}{\tesla}. The HFIPs in device D3 also disappear when warmed up to \SI{4}{\kelvin}.

The slopes of the Landau levels are used to accurately calibrate the density axis $n$ according to $n=\nu B/\phi_0$ once the filling factor $\nu$ is known. This calibration of the density is in agreement with the parallel plate estimation $n\approx C_gV_g/e$ to within \SI{10}{\percent}, where $C_g=\epsilon_\mathrm{h-BN}/t$ is the gate capacitance per unit area of h-BN, $t$ is the thickness of h-BN, and $V_g$ is the gate voltage.

\subsection{In-plane magnetotransport data for D1}

\begin{figure}[!htb]
	\includegraphics[width=\textwidth]{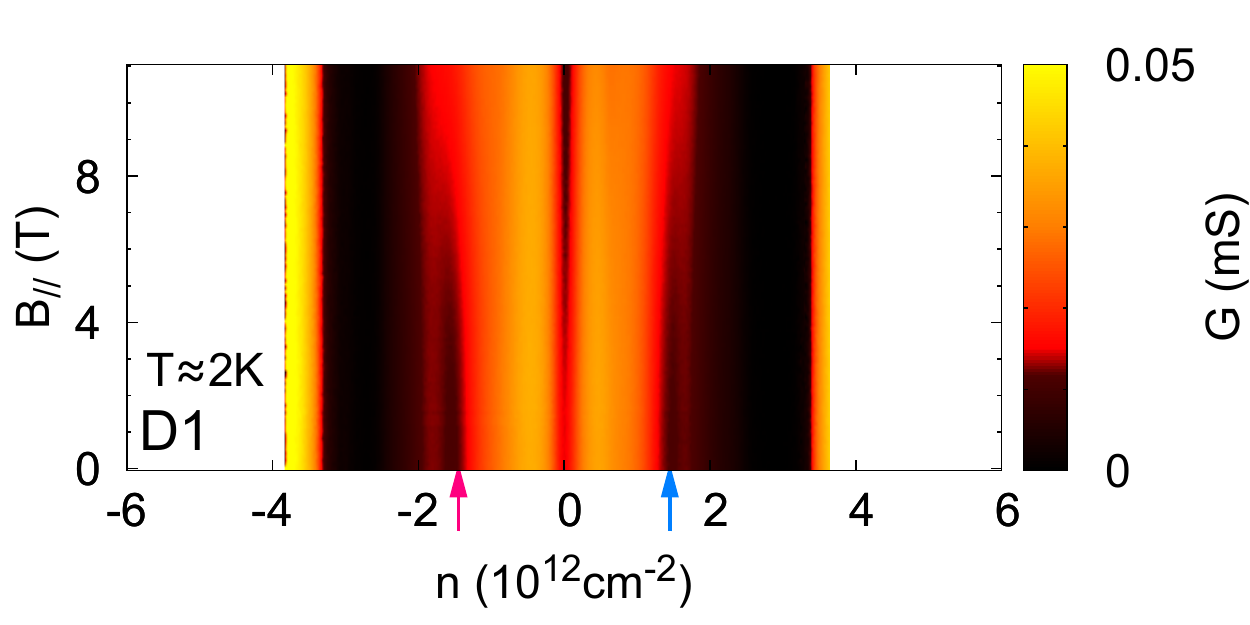}
	\caption{\label{fig:inplane}In-plane magnetic field dependence of the conductance of D1. The measurement is taken at a higher temperature of $T\approx\SI{2}{\kelvin}$. The color scale is chosen to emphasize the HFIPs. }
\end{figure}

To reveal the origin of the HFIPs and its suppression in a magnetic field, we have performed magneto-transport measurements in an in-plane configuration, \emph{i.e.} the magnetic field vector is parallel to the sample plane. Fig. \ref{fig:inplane} shows the measured conductance as a function of carrier density $n$ and in-plane magnetic field $B_\parallel$. 

The in-plane measurement is taken at a higher temperature of $T\approx\SI{2}{\kelvin}$. Combined with the degradation of the sample quality resulting from the thermal cycling that was necessary in order to change the field orientation, the HFIPs are not as well developed as in the previous measurements (Fig. \ref{fig:mt13}).  However, the gradual suppression of the HFIPs is still unambiguously observed when $B_\parallel$ is above about \SI{6}{\tesla}, slightly higher but similar to the \SIrange{4}{6}{\tesla} threshold for the perpendicular field (see Fig. \ref{fig:mt13} and main text Fig. 4a-b). Based on these observations, we conclude that the suppression of HFIPs in a magnetic field originates from the Zeeman effect on the electron spins rather than some orbital effect.

\subsection{Transport data in device D4}

\subsubsection{Four-probe measurement and $\pm3$-filling states}

Transport measurements in both D1 and D3 were performed in a two-probe configuration. Although it is generally advised to perform four-probe measurements in transport experiments, we find that the existence of multiple insulating states (both the superlattice gaps at $\pm n_s$ and HFIPs at $\pm\frac{1}{2}n_s$) frequently lead to noisy or negative $R_{xx}$ signals due to the region in the device near the voltage probes becoming insulating at a slightly different carrier density. In our case where we are mostly interested in the insulating behaviours on the order of \SI{100}{\kilo\ohm} to \SI{1}{\mega\ohm}, a contact resistance of at most a few \si{\kilo\ohm} that is typical in edge-contacted graphene device does not obscure the present data\cite{wang2013}. Thus we believe that the two-probe data presented throughout the paper is fully trustable and gives an accurate presentation of the device characteristics.

\begin{figure}[!htb]
	\includegraphics[width=\textwidth]{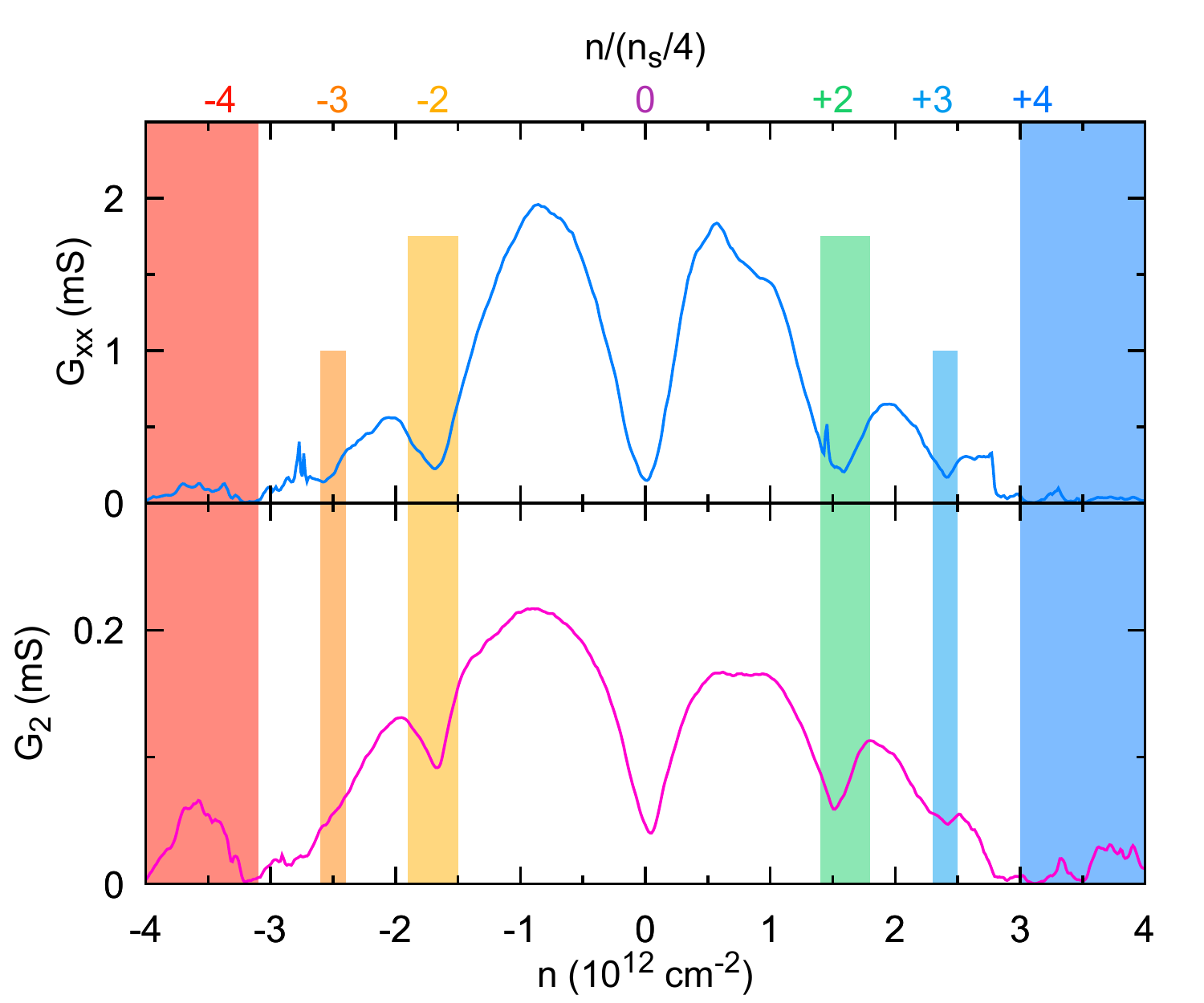}
	\caption{\label{fig:D4} Four-probe (upper panel) and two-probe (lower panel) conductance measured in device D4 at $T=\SI{0.3}{\kelvin}$. The colored vertical bars and the corresponding numbers indicate the associated integer filling inside each unit cell of the moir\'{e} pattern. Besides clear observation of the HFIPs at half-filling ($\pm$2), we also observe weak drops in the four-probe conductivity that point towards three-quarter-filling states at $\pm$3. }
\end{figure}

Here we show the measurements in a fourth device D4 which has a twist angle of $\theta=\SI{1.16+-0.02}{\degree}$. Device D4 was measured in a four-probe configuration so that the contact resistance is removed. Both the superlattice insulating states and the HFIPs do not have very high impedance in D4 (probably due to disorder and/or inhomogeneity), and therefore the previously described issues with four-probe measurements did not occur. Fig. \ref{fig:D4} shows the two-probe and four-probe conductances in device D4 measured at \SI{0.3}{\kelvin}. 

From Fig. \ref{fig:D4} it is clear that the four-probe and two-probe measurements essentially show the same features, while some weak signals appear to be better resolved in the four-probe measurements. In the four-probe data, we not only observe the HFIPs at half-filling ($\pm2$ electrons per moir\'{e} unit cell), we also see evidence for odd-filling insulating phases at $\pm3$ electrons per moir\'{e} unit cell as weak reduction in the conductance curve. Note that the existence of insulating behaviours at other integer fillings of the flat bands than $\pm2$ is a result to be expected in the Mott-like insulator picture, and further lends support to our claim that the correlated insulating behavior originates from the on-site Coulomb interaction.

\begin{figure}[!htb]
	\includegraphics[width=\textwidth]{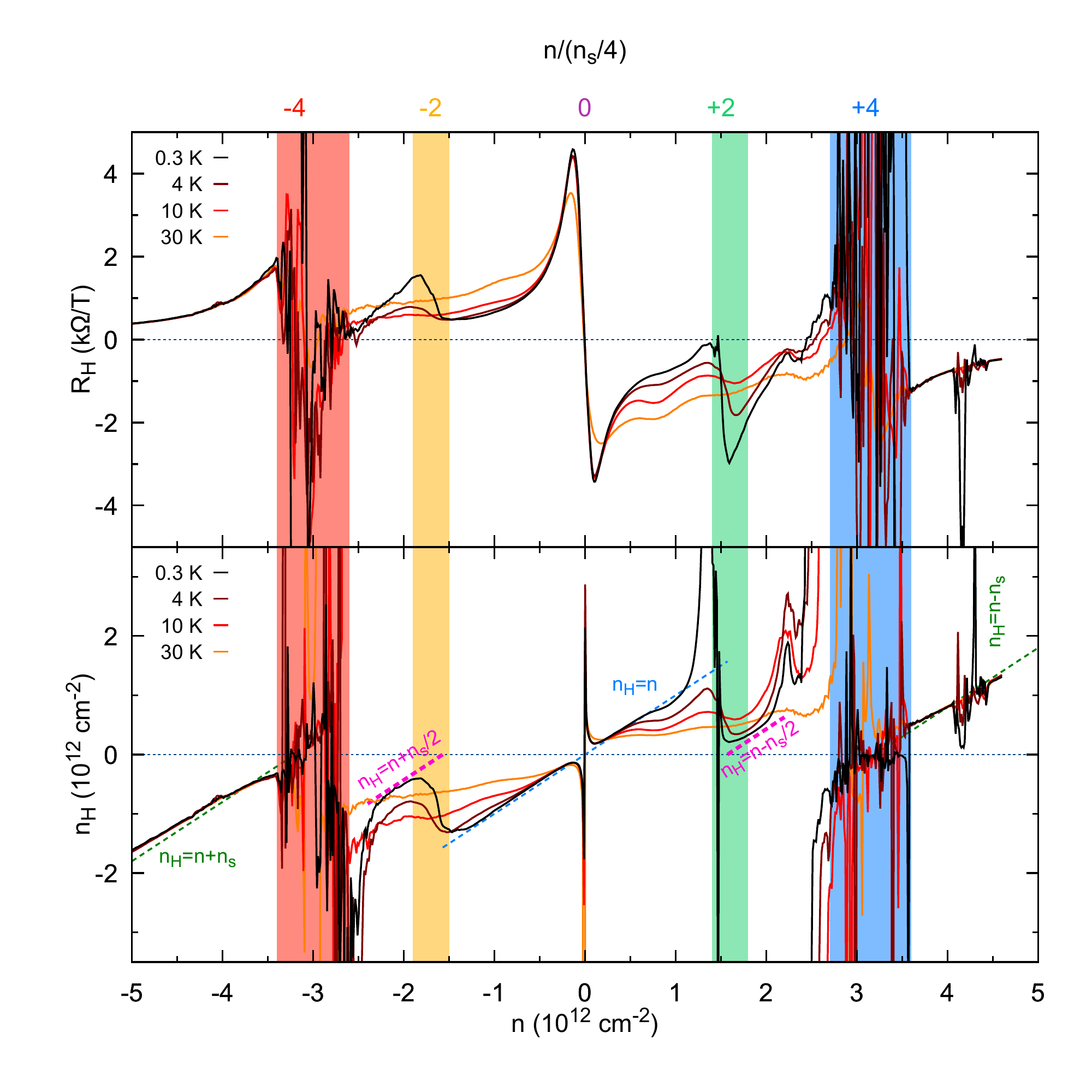}
	\caption{\label{fig:hall} Hall measurement in device D4 ($\theta=\SI{1.16}{\degree}$) at \SI{0.3}{\kelvin}, \SI{4}{\kelvin}, \SI{10}{\kelvin} and \SI{30}{\kelvin}. The upper panel shows the Hall coefficients $R_H$ and the lower panel shows the Hall density $n_H=-1/(eR_H)$. The colored vertical bars and the corresponding numbers indicate the associated integer fillings in the moir\'{e} unit cell. The x-axis is the gate-induced total charge density $n$, while the Hall density $n_H$ and its sign indicates the number density and characteristic (electron/hole) of the carriers being transported.}
\end{figure}
\subsubsection{Hall measurement}

We have also measured the device D4 in a Hall configuration ($R_{xy}$). Fig. \ref{fig:hall} shows the low-field linear Hall coefficient $R_H=R_{xy}/B$ and Hall density $n_H=-1/(eR_H)$ versus gate-induced charge density $n$. In a uniformly gated single-carrier 2D electronic gas, one expects that $n_H=n$. This is what we have measured in the density range -1.3$\sim$\SI{1.3e12}{\per\centi\meter\squared} at \SI{0.3}{\kelvin}. Near the half-filling states $n=\pm n_s/2$, however, the Hall density abruptly jumps from $n_H=n$ to a small value close to zero (but not changing its sign). Beyond half-filling, $n_H$ follows $n_H=n\pm n_s/2$, a new trend that is consistent with \emph{quasiparticles} that are generated from the half-filling states. This `resetting' effect of the Hall density gradually disappears as the temperature is raised from \SI{0.3}{\kelvin} to \SI{10}{\kelvin}, in agreement with the energy scale of the Mott-like states. At higher temperatures, the Hall density is linear with $n$ but the slope is no longer one, which might be related to the thermal energy $kT$ being close to the bandwidth, resulting in thermally excited carriers with opposite polarity reducing the net Hall effect.

We note that in good correspondence with the quantum oscillation data shown in Fig. 3b, we only see the behaviors of the new quasiparticles on one side of the Mott-like state, e.g. the side further from the charge neutrality point; between the charge neutrality point and the Mott-like state, we see an abrupt change from the typical large Fermi surface of the single-particle bands to a small Fermi surface of the new quasiparticles. This may result if the effective mass of the quasiparticles on one side of the Mott-like gap is considerably greater than the other side, so that the oscillation and Hall effect become difficult to observe very close to the metal-insulator transition.

\subsection{Determination of Twist Angles}
Accurate determination of the twist angles of the samples is of utmost importance in understanding the magic-angle physics. We use two independent methods to determine the twist angle from transport data.

\subsubsection{Superlattice density}

The superlattice density $n_s$, defined by the density required to fill one band in the superlattice, is related to the twist angle by 
\begin{equation}
\label{eq:ns}n_s = \frac{4}{A} \approx \frac{8 \theta^2}{\sqrt{3}a^2},
\end{equation}
where $A$ is the unit cell area, $a=\SI{0.246}{\nano\meter}$ is the lattice constant of graphene.

\begin{figure}[!htb]
	\includegraphics[width=\textwidth]{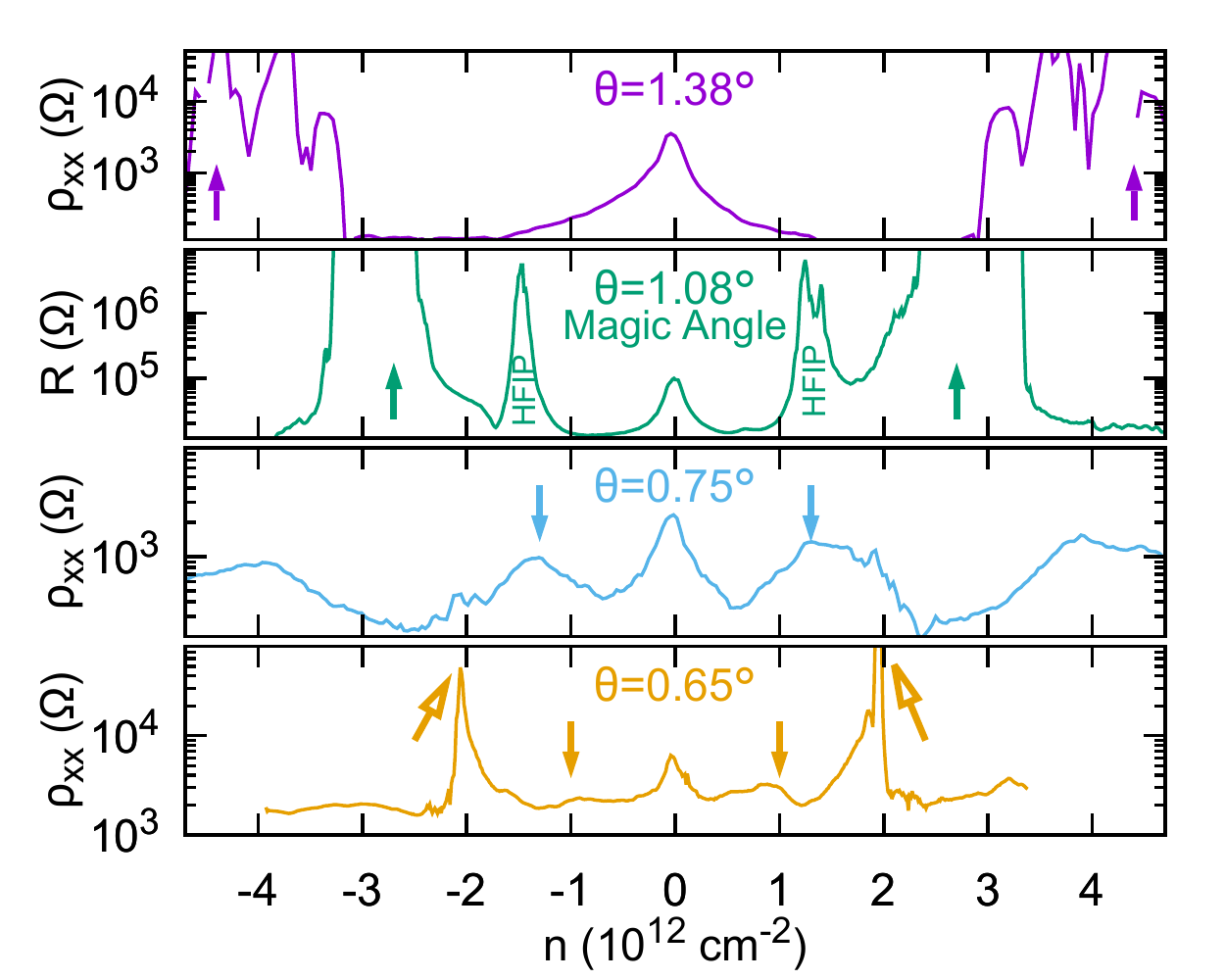}
	\caption{\label{fig:angles} Resistivity (resistance for $\theta=\SI{1.08}{\degree}$) measurements for four samples with different twist angles, $\theta=1.38, 1.08, 0.75$ and \SI{0.65}{\degree}. The solid arrows point towards superlattice features at $\pm n_s$, and empty arrows point to $\pm 2n_s$. HFIPs have only been observed in first-magic-angle samples so far.}
\end{figure}

At approximately $\SI{1}{\degree}<\theta<\SI{3}{\degree}$, the superlattice densities $\pm n_s$ are associated with a pair of single-particle bandgaps at their corresponding Fermi energy \cite{moon2012, cao2016}. Therefore, the measured density of the superlattice insulating states can be used to directly estimate $\theta$ according to Eq. (\ref{eq:ns}). Due to localized states, the accurate value of $n_s$ is difficult to pinpoint at zero magnetic field, and the estimated $\theta$ is accurate to about \SIrange{0.1}{0.2}{\degree}. Fig. \ref{fig:angles} shows the resistivity (resistance for magic-angle device D1) for four different samples of twist angles 
$\theta=1.38, 1.08, 0.75, \SI{0.65}{\degree}$ respectively. At $\theta=1.38, 1.08\si{\degree}$, the positions of the superlattice gaps clearly provide an estimation of $\theta$.

However, it is noted in Ref.\cite{kim2017} that the apparent resistance peaks in the transport data may not correspond to $n_s$ but instead $2n_s$, once the twist angle is below \SI{1}{\degree}. We have observed a similar phenomenon when twist angle is as small as \SI{0.65}{\degree}. This complicates the determination of twist angles, since one encounters an ambiguity of whether the feature one observes corresponds to $n_s$ or $2n_s$, which can result in the twist angle wrong by a factor of $\sqrt{2}$.

\begin{figure}[!htb]
	\includegraphics[width=\textwidth]{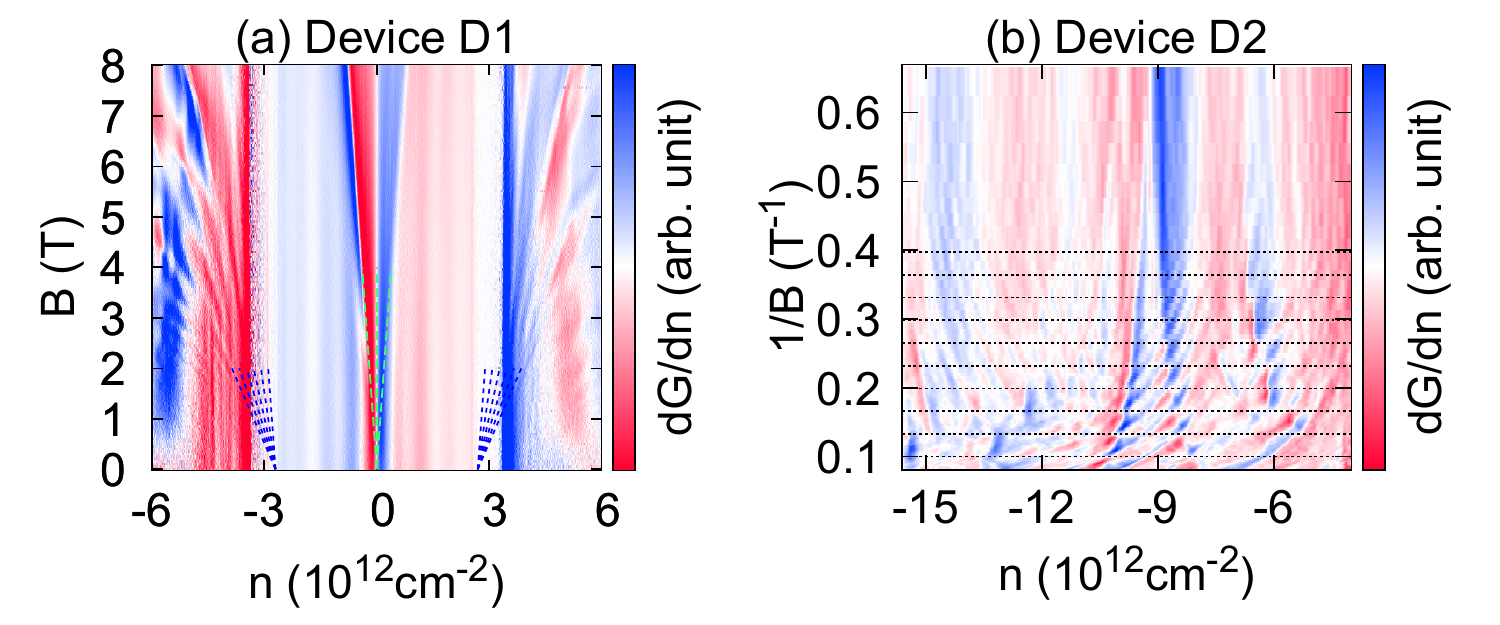}
	\caption{\label{fig:s1s2} (a) Derivative of magneto-conductance data of device D1, measured at \SI{4}{\kelvin}. The dashed fans labels the main (green) and satellite (blue) Landau fans respectively. (b) The derivative of magneto-conductance data of device D2 plotted as a function of $1/B$. The horizontal lines have a uniform spacing of \SI{0.033}{\per\tesla}. }
\end{figure}

To more accurately determine the twist angle and avoid this ambiguity, we use the fact that each band edge of the miniband structure has its own Landau levels \cite{moon2012, cao2016, kim2016}. Fig. \ref{fig:s1s2}(a) shows the derivative of magneto-conductance data of device D1. The Landau levels emanating from $n_s=\SI{2.7+-0.1e12}{\per\centi\meter\squared}$ can be clearly seen, which translates to $\theta=\SI{1.08+-0.02}{\degree}$ according to Eq. (\ref{eq:ns}). Since the intersection points of the Landau levels can be determined relatively accurately (uncertainty of about \SI{1e11}{\per\centi\meter\squared}), the twist angle can be determined with an uncertainty of about \SI{0.02}{\degree} near the first magic angle.

\subsubsection{Hofstadter's Oscillation}

The effect of applying strong magnetic fields such that the magnetic length becomes comparable with the unit cell size is well described by Hofstadter's butterfly model \cite{hofstadter1976}. In density space, this model is better captured in Wannier's picture \cite{wannier1978}. In the Wannier diagram, all Landau levels are described by $n/n_s = \phi/\phi_0 + s$, where $\phi$ is the magnetic flux through a unit cell. $s=0$ labels the main Landau fan and $s=\pm1$ is the (first) satellite fan, etc. Adjacent Landau fans intersect when $\phi/\phi_0=1/q$ or equivalently $1/B=qA/\phi_0$,  where $q$ is an integer. Therefore, in the experiments one would expect to see Landau level crossings at periodic intervals of $1/B$, where the periodicity is proportional to the unit cell area $A$. This effect has been observed in other 2D superlattice systems, and can be utilized to cross-check the twist angles extracted from other methods \cite{hunt2013,ponomarenko2013,dean2013}.

Fig. \ref{fig:s1s2}(b) shows the magneto-transport data (first derivative with respect to density) of device D2 at higher doping densities, plotted versus $n$ and $1/B$. A periodic crossing of Landau levels is clearly observed near \SI{-9e12}{\per\centi\meter\squared}. The periodicity is \SI{0.033+-0.001}{\per\tesla}, which gives $A=\SI{1.37+-0.04e-12}{\centi\meter\squared}$ and $\theta=\SI{1.12+-0.01}{\degree}$, compared to $\theta=\SI{1.12+-0.02}{\degree}$ extracted using the previous method ($n_s=\SI{2.9+-0.1e12}{\per\centi\meter\squared}$).

\section{Capacitance measurement}
\subsection{General Description}

\begin{figure}[!htb]
	\includegraphics[width=0.7\textwidth]{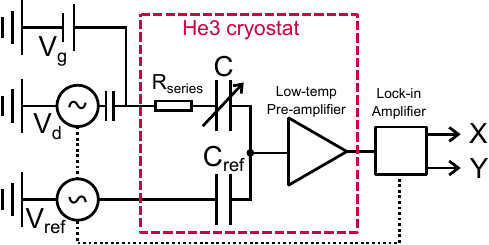}
	\caption{\label{fig:cap} Schematic for the low-temperature capacitance bridge. The X and Y outputs from the lock-in amplifier refer to the in-phase and out-of-phase components respectively. $C$ and $R$ are the capacitance and resistance of the sample. $V_g$ is the DC gate voltage. All connections into and out from the cryostat are made with coaxial cables.}
\end{figure}

In order to measure a tiny capacitance change between the TwBLG device to the metal gate, we use a balanced capacitance bridge as illustrated in Fig. \ref{fig:cap} \cite{ashoori1992}.

The capacitance bridge is first `balanced' by adjusting the ratio $V_d/V_\mathrm{ref}$ such that the output from the pre-amplifier is close to zero. When balanced, any small change of the device capacitance $C$ is linearly proportional to the output signal. The reference capacitance $C_\mathrm{ref}$ used in our experiment is approximately \SI{40}{\femto\farad}, and the device geometrical capacitance is approximately \SI{7}{\femto\farad}. The ac excitation used in our measurements is \SI{3}{\milli
\volt} at $f=\SI{150}{\kilo\hertz}$.

The in-phase and out-of-phase components of the measured signal are (to the leading order) proportional to the change in capacitance $\Delta C/C_\mathrm{ref}$ and dissipation $\omega C R_\mathrm{series}$, respectively ($\omega=2\pi f$). Therefore from the out-of-phase component, one can in principle extract the effective resistance $R_\mathrm{series}$ in series with the device capacitance $C$. In reality, due to the complications that the actual resistance is distributed across the device, and that the device capacitance only constitutes about 1/5 of the total measured capacitance (the remainder being stray capacitance from bonding pads and bonding wires), the dissipation data should be interpreted as a qualitative measure of the device resistance. %In Fig. 3a in the main text, $R_\mathrm{series}$ is approximated as
%\[R_\mathrm{series} = \frac{Y}{\omega C_g},\]
%where $C_g\approx\SI{7.5}{\femto\farad}$ is the DC gating capacitance, calibrated with the slopes of Landau levels (see Sec. \ref{sec:ll}).

\subsection{Estimation of the Fermi velocity}

\begin{figure}[!htb]
	\includegraphics[width=\textwidth]{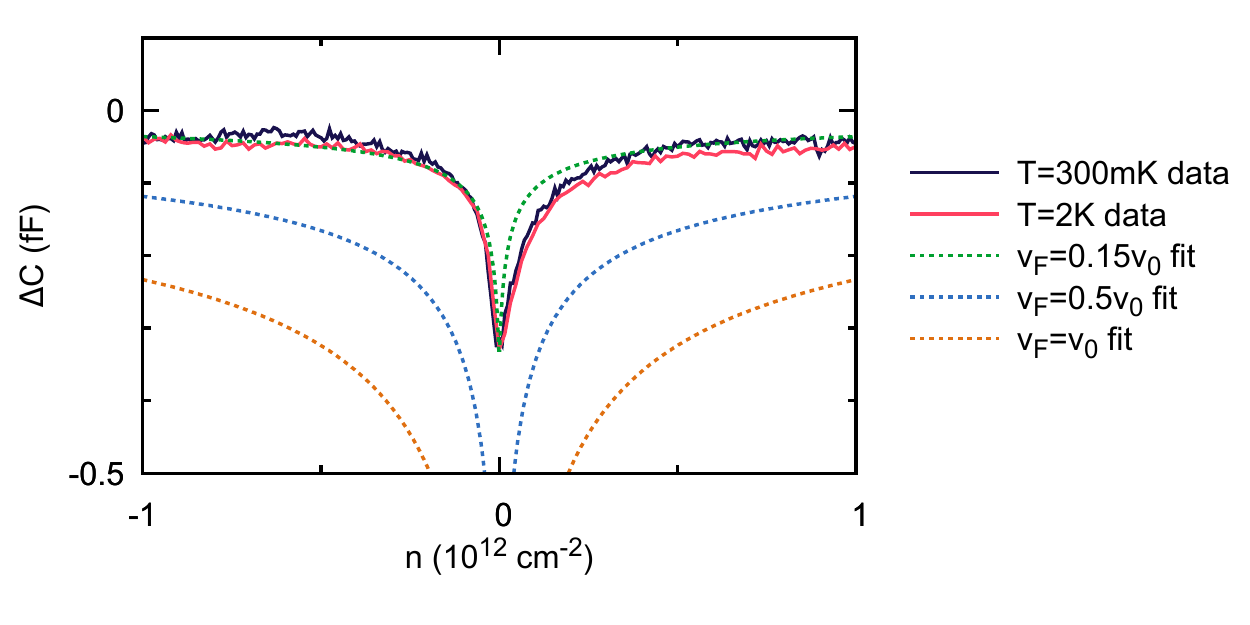}
	\caption{\label{fig:capvf} Capacitance of device D2 near the charge neutrality point, and fitting curves according to Eq. (\ref{eq:ctot}) with different Fermi velocities. $v_0=\SI{1e6}{\meter\per\second}$ is the Fermi velocity in pristine graphene.}
\end{figure}

The measured capacitance is the series sum of geometric capacitance $C_\mathrm{geom}$ and quantum capacitance $C_q$. The latter is directly proportional to the density of states (DOS) in TwBLG. Therefore, by analyzing the quantum capacitance $C_q$ as a function of carrier density $n$, one can extract the dependence of DOS on $n$, and subsequently deduce the Fermi velocity.

In the zero temperature limit, the quantum capacitance is related to the DOS by $C_q = e^2 D(E_F)$, where $E_F$ is the Fermi energy. A model system for TwBLG near the charge neutrality consists of massless Dirac fermions with Fermi velocity $v_F$ and 8-fold degeneracy (spin, valley, layer), the DOS is \cite{xia2009, fang2007, wallace1947}
\begin{equation}
D(E_F) = \frac{4}{\pi} \frac{E_F}{(\hbar v_F)^2}.
\end{equation}

Since $E_F=\hbar v_F k_F$ is related to the density $n$ by
\begin{align}
n = 8\cdot\frac{1}{(2\pi)^2}\cdot \pi k_F^2 = \frac{2}{\pi} \frac{E_F^2}{(\hbar v_F)^2}, \\
E_F = \hbar v_F \sqrt{\frac{n\pi}{2}},
\end{align}
where the factor 8 comes from spin, valley, and layer, the quantum capacitance of the TwBLG is therefore written as
\begin{equation}
\label{eq:cq}C_q = e^2 \frac{2\sqrt{2}}{\sqrt{\pi} \hbar v_F} \sqrt{|n| + n_d}.
\end{equation}

Due to disorder, the spatially averaged DOS at the Dirac point ($n=E_F=0$) will not be absolutely zero. Therefore a phenomenological $n_d\sim\SI{1e10}{\per\centi\meter\squared}$ is added in the expression above \cite{xia2009}.

The measured capacitance is then
\begin{equation}
\label{eq:ctot}\frac{1}{C} = \frac{1}{C_\mathrm{geom}} + \frac{1}{C_q}.
\end{equation}

In Fig. \ref{fig:capvf}, we show the measured capacitance near the Dirac point and fitting curves according to Eq. (\ref{eq:ctot}) and Eq. (\ref{eq:cq}). The $C_\mathrm{geom}$ is again approximated by the DC gating capacitance $C_g \approx \SI{7.5}{\femto\farad}$. We find that using parameters $v_F = \SI{0.15e6}{\meter\per\second}$ and $n_d=\SI{1.0e10}{\per\centi\meter\squared}$ gives a reasonable fit to the data measured at both $T=\SI{300}{\milli\kelvin}$ and $T\approx\SI{2}{\kelvin}$. 

Finally, we note that the fitting for $v_F$ is sensitive to the value used for $C_\mathrm{geom}$. For example, using a $C_\mathrm{geom}$ value \SI{30}{\percent} larger than the value we used above, we find a Fermi velocity of $v_F=\SI{0.1e6}{\meter\per\second}$. Similarly, using a value \SI{15}{\percent} smaller than the said value we find $v_F=\SI{0.2e6}{\meter\per\second}$. Nonetheless, the analysis present here suffices to demonstrate that the Fermi velocity is indeed greatly reduced in the capacitance device D2. The slightly larger Fermi velocity compared to that measured in the transport device D1 $v_F=\SI{0.04e6}{\meter\per\second}$ can be attributed to the slightly larger twist angle of device D2 $\theta=\SI{1.12}{\degree}$, which is further from the first magic angle $\theta_\mathrm{magic}^{(1)}\approx \SI{1.05}{\degree}$ than device D1 $\theta=\SI{1.08}{\degree}$.

\section{Error bar in Fig. 2a Inset in Main Text}

The error bars in Fig. 2a are deduced using the following criteria: 

\begin{enumerate}
\item For the transport devices D1, D3 and D4, the endpoints of the error bars correspond to the points where the conductance rises to \SI{10}{\percent} of the peak value on that side.
\item For the capacitance device D2, since the peaks are very sharp (see Fig. 3a in main text), the error bar corresponds to the entire width of peaks in the $R_\mathrm{series}$ data.
\end{enumerate}

\section{Quantum Oscillations and Extraction of $m^*$}

\begin{figure}[!htb]
	\includegraphics[width=0.7\textwidth]{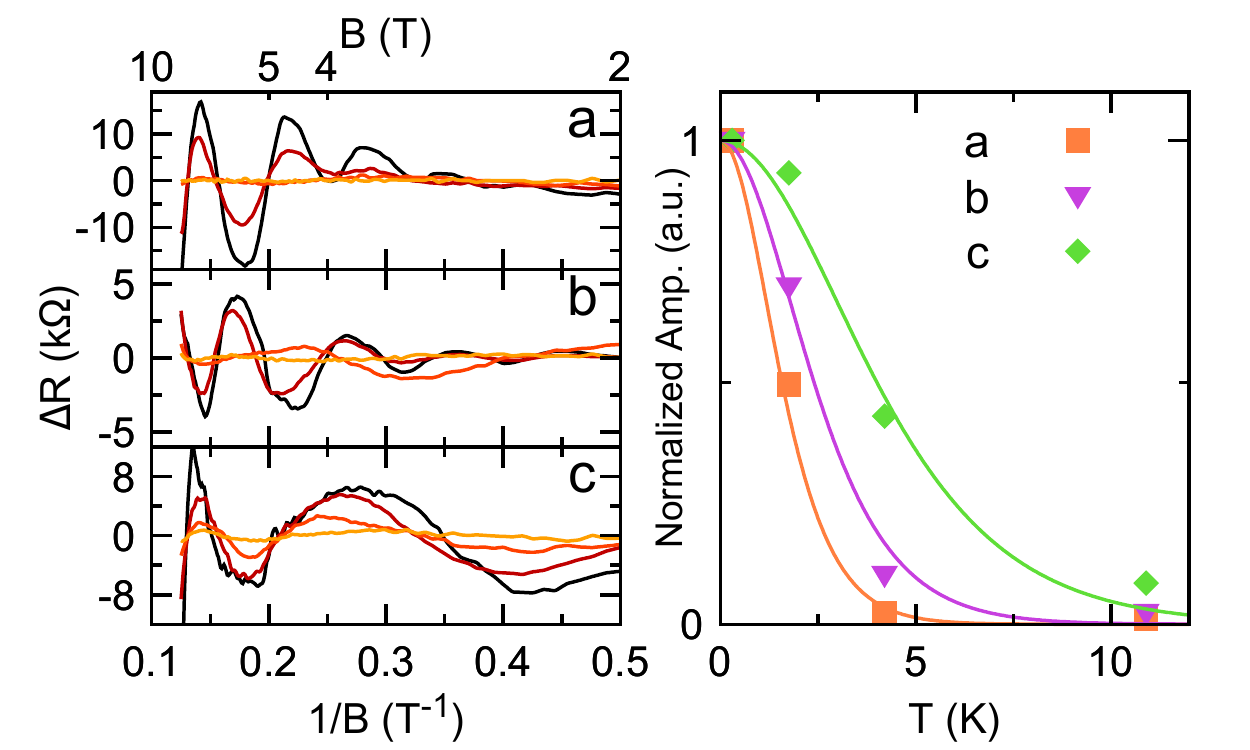}
	\caption{\label{fig:sdh} Temperature-dependent magneto-conductance of device D1 at gate voltage (a) $V_g=\SI{-2.28}{\volt}$, (b) $V_g=\SI{-0.68}{\volt}$ and (c) $V_g=\SI{1.08}{\volt}$. The temperatures are from dark to bright, \SI{0.3}{\kelvin}, \SI{1.7}{\kelvin}, \SI{4.2}{\kelvin} and \SI{10.7}{\kelvin} respectively. The figure on the right summarizes the oscillation amplitudes of the most prominent peaks in (a-c). The curves are fitted according to the L-K formula Eq. (\ref{eq:lk}).}
\end{figure}

We performed magnetotransport measurements from \SI{0.3}{\kelvin} to \SI{10}{\kelvin}. At each gate voltage, a polynomial background of resistance in $B$ is first removed, and then the oscillation frequency and the effective mass is analyzed. Examples of the SdH oscillations and their temperature dependences at a few representative gate voltages are shown in Fig. \ref{fig:sdh}. Temperature dependence of the most prominent peak is fitted with the Lifshitz-Kosevich formula applied to conductance
\begin{equation}
\label{eq:lk}\Delta R \propto \frac{\chi}{\sinh(\chi)}, \quad \chi=\frac{2\pi^2 k T m^*}{\hbar e B},
\end{equation}
and the cyclotron mass $m^*$ is extracted from the fitting. 

Within the flat bands, the quantum oscillations universally disappear at \SI{10}{\kelvin} except very close to the Dirac point, again consistent with the large electron mass and greatly reduced Fermi velocity near the first magic angle. 

\begin{figure}[!htb]
	\includegraphics[width=\textwidth]{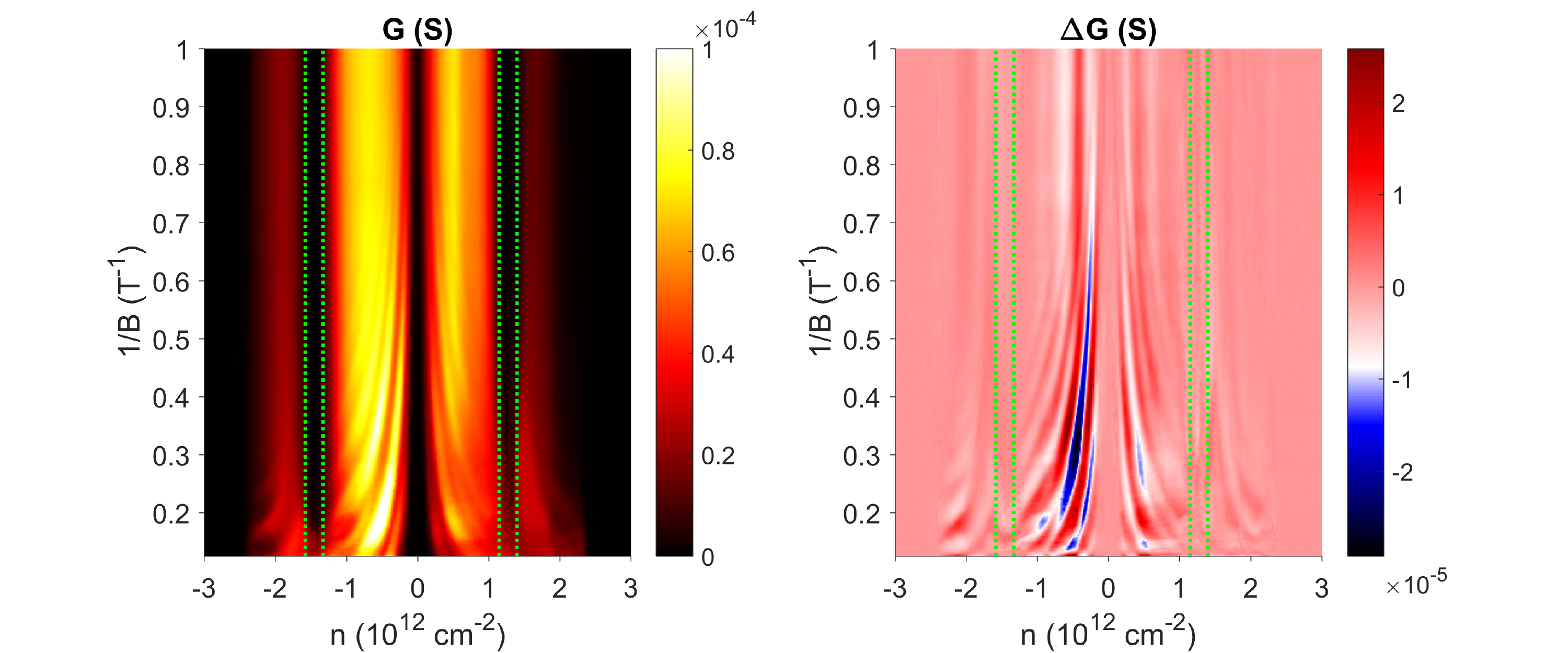}
	\caption{\label{fig:invB} Quantum oscillations in device S1 with twist angle of \SI{1.08}{\degree}. Left: raw experimental data plotted versus $n$ and $1/B$. Right: same data with a polynomial background in $B$ removed for each density. The green boxes denote the range of density for the half-filling insulating states.}
\end{figure}

The quantum oscillations in device D1 is shown in Fig. \ref{fig:mt13}(a). At a first glance, it may seem that the Landau levels emanating from the Dirac point `penetrate' the half-filling Mott-like states and continue towards the band edges. However, upon closer examination this is not the case. Fig. \ref{fig:invB} shows the same data as in Fig. \ref{fig:s1s2}a, but plotted versus $1/B$ instead of $B$. Here it can be seen that at densities beyond the half-filling states the oscillations are clearly not converging at the Dirac point, but instead the half-filling states themselves. The oscillation frequencies extracted from this data are plotted in Fig. 3b.

\section{Some theoretical and numerical insights in T\MakeLowercase{w}BLG near Magic Angles}

\subsection{Band Structure near the Magic Angles}
\label{sec:band}

\begin{figure}[!htb]
	\includegraphics[width=\textwidth]{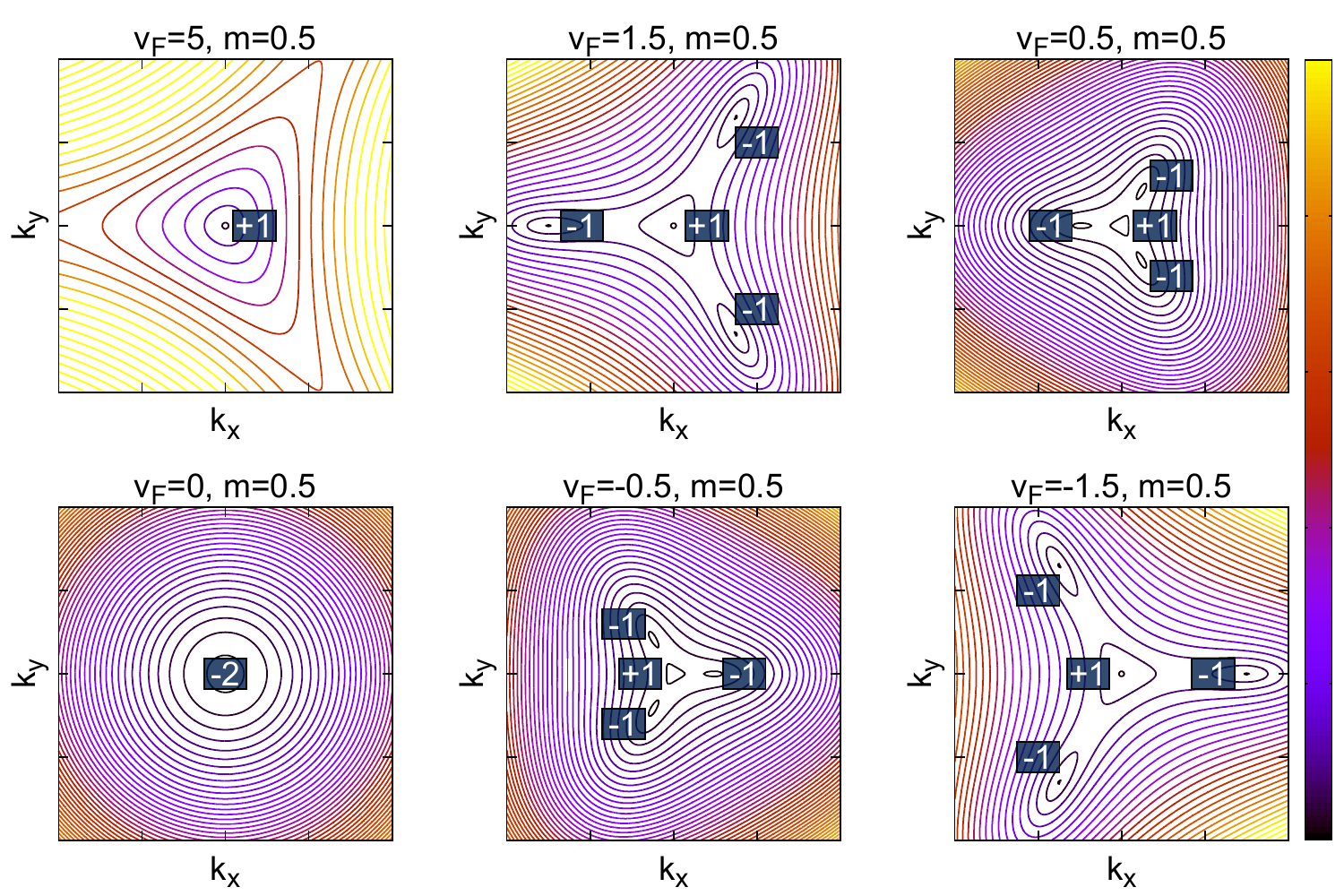}
	\caption{\label{fig:hamband} $E_+$ dispersion from Eq. (\ref{eq:hamdis}) for different $v_F$ and $m=0.5$. The $k_x$ and $k_y$ scale is $[-2,2]$, the colorscale for the energy axis is $[0,10]$. The associated winding number of each band touching point is labeled in the figure.}
\end{figure}

The general evolution of the band structure of TwBLG above the first magic angle is described in a number of earlier works \cite{morell2010, bistritzer2011, moon2012, laissardiere2012, santos2012, fang2016}. The low-energy band structure consists of two Dirac cones (each is 4-fold degenerate due to valley and spin), with a renormalized Fermi velocity
\begin{equation}
\label{eq:vf}v_F(\theta) = \frac{1-3\alpha^2}{1+6\alpha^2} \approx 1-9\alpha^2 (\alpha \le 1),
\end{equation}
where $\alpha = w/v_0 k_\theta$ is the dimensionless interlayer hopping amplitude \cite{bistritzer2011}($w$, $v_0$ are the interlayer hopping energy and original Fermi velocity in graphene, $k_\theta\approx K \theta$ is the interlayer momentum difference). $v_F(\theta)$ passes through zero at $\alpha=1/\sqrt{3}$, which defines the first magic angle $\theta_\mathrm{magic}^{(1)}$. To the best of our knowledge the detailed evolution of the band structure near the magic angles has not been addressed in the literature. Specifically, we ask the following question: as the Fermi velocity at the Dirac points changes sign, how does the associated \emph{winding number} evolve? Close to a Dirac point, the effective two-band Hamiltonian can be written as
\begin{equation}
\mathcal{H}(k) = v_F(\theta)\sigma\cdot k + \mathcal{O}(k^2) = \begin{pmatrix}
\mathcal{O}(k^2) & v_F(\theta) \bm{k^\dagger} + \mathcal{O}(k^2) \\
v_F(\theta) \bm{k} + \mathcal{O}(k^{2}) & \mathcal{O}(k^2)
\end{pmatrix},
\end{equation}
in which $\bm{k}=k_x + ik_y$. When $v_F(\theta)\rightarrow 0$ near the first magic angle, the terms linear in $k$ vanish and the dispersion is dominated by the next-leading-order $k^2$ terms. A simple form of the $k^2$ term is
\begin{equation}
\label{eq:ham}
\mathcal{H}(k) = \begin{pmatrix}
0 & v_F(\theta) \bm{k^\dagger} + \frac{1}{2m} \bm{k}^2 \\
v_F(\theta) \bm{k} + \frac{1}{2m} \bm{k}^{\dagger2} & 0
\end{pmatrix},
\end{equation}
$m$ is a parameter with the dimension of mass. In fact, this Hamiltonian describes the low-energy band dispersion of monolayer graphene with third-nearest-neighbor hopping \cite{bena2011, montambaux2012}, as well as bilayer graphene with Bernal stacking and trigonal warping \cite{montambaux2012, mccann2013,mccann2006}. The eigenvalues of this Hamiltonian are 
\begin{equation}
\label{eq:hamdis}E_\pm(k) = \pm \sqrt{[v_F k_x + \frac{1}{2m}(k_x^2 - k_y^2)]^2 + [v_F k_y - \frac{1}{m}k_xk_y]^2}.
\end{equation}
The evolution of the dispersion described by Eq. (\ref{eq:hamdis}) with varying $v_F$ and constant $m=0.5$ is shown in Fig. \ref{fig:hamband}. The winding number associated with a Dirac point is defined by 
\begin{equation}
w = \frac{i}{2\pi}\oint_C \left<k\right|\nabla_k\left|k\right>\cdot dk,
\end{equation}
where $C$ is a loop around that Dirac point. The winding number follows a conservation law when the motion and merging of Dirac points are considered \cite{goerbig2017}. The winding number of each band touching point is labeled in Fig. \ref{fig:hamband}. 

When $v_F\rightarrow 0$ there exist three \emph{additional} Dirac points with \emph{opposite} winding numbers ($-1$) to the main Dirac point ($+1$). Therefore at $v_F=0$ when all four Dirac points merge, the winding number is $-2$, since the total winding number cannot change. %This explains why the Landau levels that we observed in the magic-angle devices have a bilayer-graphene-like sequence of $\pm(4, 8, 12, \ldots)$ (see Sec. \ref{sec:ll}).

\begin{figure}[!htb]
	\includegraphics[width=\textwidth]{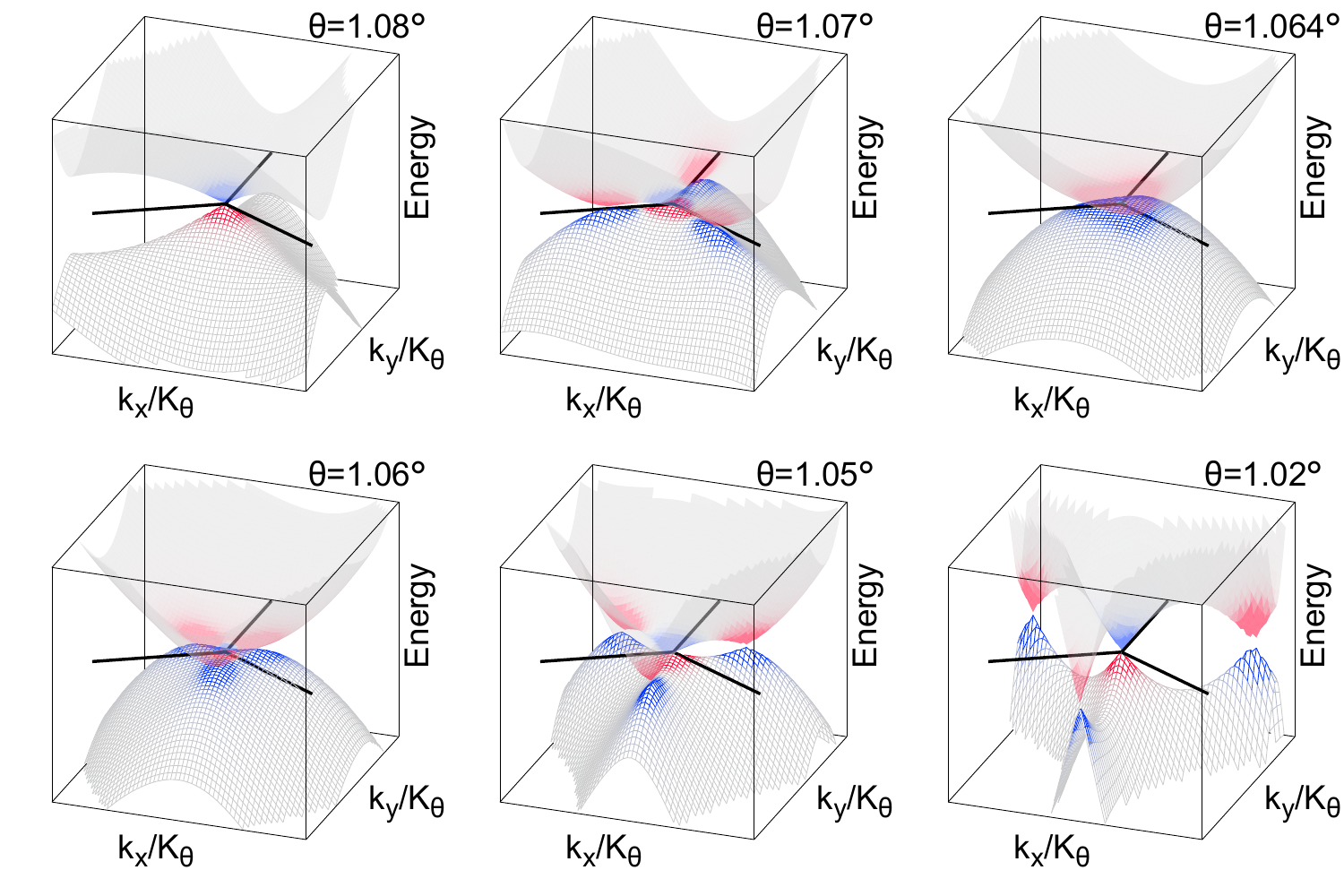}
	\caption{\label{fig:macdonald} The evolution of the low-energy band structure of TwBLG near the first magic angle $\theta_\mathrm{magic}^{(1)}=\SI{1.064}{\degree}$. The coloring shows the hotspots of the Berry curvature at each band touching point. The energy axis spans an extremely small range of \SIrange{-50}{50}{\micro\electronvolt}. The momentum axes are measured by $K_\theta\approx K\theta$ and the range for both $k_x/K_\theta$ and $k_y/K_\theta$ is $[-0.1,0.1]$. The center of the momentum space is the $K_s$ point of the MBZ, and the thick lines denotes the $K_s$ -- $M_s$ -- $K^\prime_s$ direction. All results are shown for the $K$-valley continuum description of TwBLG \cite{bistritzer2011}.}
\end{figure}

The simple Hamiltonian form of Eq. (\ref{eq:ham}) is an educated guess. We performed numerical calculations of the winding number using the continuum model for TwBLG \cite{bistritzer2011, santos2012} and the numerical method in Ref. \cite{fukui2005}. The results are summarized in Fig. \ref{fig:macdonald}. We find that near the first magic angle, $\theta_\mathrm{magic}^{(1)}=\SI{1.064}{\degree}$, the picture described in Fig. \ref{fig:hamband} is exactly what happens at each corner of the mini Brillouin zone (MBZ). The complication that arises when one considers the entire MBZ is that, for a given valley (\emph{e.g.} $K$), the two inequivalent corners of the MBZ have the \emph{same} winding number, because they are the hybridized result of the same valley ($K$) of opposite \emph{layers} (see Fig. 1d of the main text). Global time reversal symmetry is preserved by mapping to the other valley ($K^\prime$). Therefore, for a given valley $K$, when the twist angle is reduced from large angles where the winding numbers of the two corners are (+1, +1) to the first magic angle where the winding numbers are (-2,-2), a net winding number change $\Delta w = 6$ occurs between the two lowest energy bands. Further theoretical work is necessary to elucidate the physics behind this winding number evolution near the first magic angle.

In summary, we have shown that at \emph{exactly} the first magic angle, the Dirac point at each corner of the MBZ ($K_s$ and $K_s^\prime$) becomes \textbf{a parabolic band touching with winding number -2}, similar to bilayer graphene with Bernal stacking except that the two corners have the \emph{same} winding number. The calculation corresponding to the first magic angle in Fig. \ref{fig:macdonald} can be fit to a paraboloid, giving an effective mass of $m=1.1m_e$. This value can be viewed as the asymptotic limit of the effective mass near the charge neutrality point as $v_F\rightarrow0$ ($\theta\rightarrow \theta_\mathrm{magic}^{(1)}$). 

\begin{figure}[!htb]
	\includegraphics[width=\textwidth]{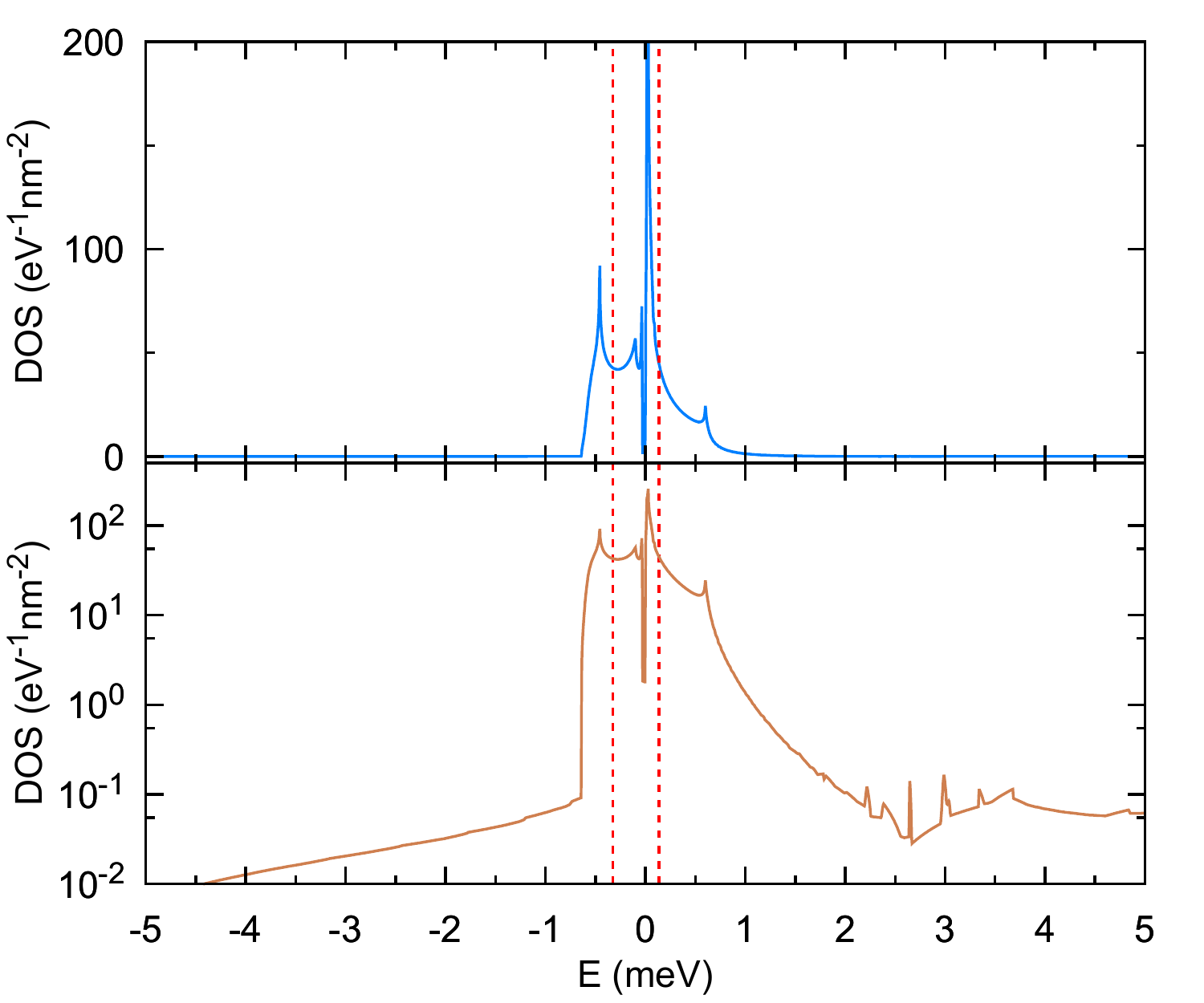}
	\caption{\label{fig:dos} Single-particle DOS in TwBLG at $\theta=\SI{1.08}{\degree}$, in linear (top) and logarithmic scale (bottom). The red dashed lines denotes the energy where the lower and upper flat bands are half-filled, respectively. The results are numerically obtained using the continuum model described in \cite{bistritzer2011}.}
\end{figure}

\subsection{Density of States (DOS) in Magic-Angle T\MakeLowercase{w}BLG}

Despite our simplistic representation of the DOS in the flat-bands of magic-angle TwBLG in Fig. 4d-f, the actual single-particle DOS profile of TwBLG is rather complex with multiple van Hove singularities (vHs). Here we show a DOS versus energy plot calculated with the continuum model as presented in \cite{bistritzer2011} for $\theta=\SI{1.08}{\degree}$.

\end{document}